\documentclass{article}

\usepackage{microtype}
\usepackage{graphicx}
\usepackage{booktabs} % for professional tables

\usepackage{hyperref}

\usepackage[accepted]{mlsys2025}
\usepackage{amsmath}
\usepackage{amssymb}
\usepackage{amsfonts}
\usepackage{amsthm}
\usepackage{todonotes}
\usepackage{graphicx}
\usepackage{subcaption} % for subfigures
\usepackage{xspace}
\usepackage{multirow}
\usepackage{cleveref}
\usepackage{enumitem}
\usepackage{pifont}
\usepackage{makecell}

\newcommand{\outline}[1]{\noindent \textbf{\textcolor{red}{[outline: #1]}}}

\newcommand{\fuwei}[1]{\textbf{\textcolor{purple}{#1}}}
\newcommand{\sysname}{\textsc{ReaL}\xspace}

\newcommand{\hz}[1]{\textcolor{blue}{[HZ: #1]}}

\newcommand{\arxiv}[1]{{#1}}

\mlsystitlerunning{\sysname: Efficient RLHF Training of Large Language Models with Parameter Reallocation}

\begin{document}

\twocolumn[
\mlsystitle{\sysname: Efficient RLHF Training of Large Language Models with Parameter Reallocation}

% unless you've provided the [accepted] option to the mlsys2024


\mlsyssetsymbol{equal}{*}

\begin{mlsysauthorlist}
\mlsysauthor{Zhiyu Mei}{equal,thu,sqz}
\mlsysauthor{Wei Fu}{equal,thu,sqz}
\mlsysauthor{Kaiwei Li}{id}
\mlsysauthor{Guangju Wang}{openpsi}
\mlsysauthor{Huanchen Zhang}{thu,sqz}
\mlsysauthor{Yi Wu}{thu,sqz,openpsi}
\end{mlsysauthorlist}

\mlsysaffiliation{thu}{Institute for Interdisciplinary Information Science, Tsinghua University, Beijing, China}
\mlsysaffiliation{sqz}{Shanghai Qi Zhi Institute, Shanghai, China}
\mlsysaffiliation{openpsi}{OpenPsi Inc.}
\mlsysaffiliation{id}{Independent Researcher}

\mlsyscorrespondingauthor{Zhiyu Mei}{meizy20@mails.tsinghua.edu.cn}
\mlsyscorrespondingauthor{Wei Fu}{fuwth17@gmail.com}
\mlsyscorrespondingauthor{Yi Wu}{jxwuyi@gmail.com}

\mlsyskeywords{LLM, RLHF, distributed training}

\vskip 0.3in

\begin{abstract}
Reinforcement Learning from Human Feedback (RLHF) is a pivotal technique for empowering large language model (LLM) applications. 
Compared with the supervised training process of LLMs, the RLHF training process is much more sophisticated, requiring a diverse range of computation workloads with intricate dependencies between multiple LLM instances. 
Therefore, simply adopting the fixed parallelization strategies from supervised training for LLMs can be insufficient for RLHF and result in low training efficiency.
To overcome this limitation, we propose a novel technique named \emph{parameter \underline{\textsc{ReaL}}location}, which 
dynamically adapts the parallelization strategies for different workloads during training by redistributing LLM parameters across the training cluster.
Building upon this idea, we introduce \sysname, a pioneering system for efficient RLHF training.
\sysname introduces the concept of an \textit{execution plan}, which defines 
a fine-grained resource allocation and parallelization strategy particularly designed for RLHF training.
Based on this concept, \sysname employs a tailored search algorithm with a lightweight run-time estimator to automatically discover an efficient execution plan for an instance of RLHF experiment.
Subsequently, the runtime engine deploys the selected plan by effectively parallelizing computations and redistributing parameters.
We evaluate \sysname on the LLaMA models with up to 70 billion parameters and 128 GPUs.
The experimental results demonstrate that \sysname achieves speedups of up to 3.58× compared to baseline methods.
Furthermore, the execution plans generated by \sysname exhibit an average of $81\%$ performance improvement over heuristic approaches based on Megatron-LM in the long-context scenario.
The source code of \sysname is publicly available at~\url{https://github.com/openpsi-project/ReaLHF}.
\end{abstract}
]



\printAffiliationsAndNotice{\mlsysEqualContribution} % otherwise use the standard text.

\section{Introduction}

Large Language Models (LLMs) such as ChatGPT~\cite{chatgpt} have amazed the world with their powerful capabilities. Their success relies on the enormous model sizes, e.g., GPT-3~\cite{gpt3} has 175 billion parameters. 
% Because each graphic processing unit (GPU) has limited memory, to perform supervised training for such an expansive model, the computation along with the model parameters must be distributed across vast GPU clusters~\cite{megascale,megatron,megatron2}. 
Because each graphic processing unit (GPU) has limited memory, to train such an expansive model, the computation along with the model parameters must be distributed across a vast GPU cluster. 
Recent literature has proposed a wide range of parallelization strategies~\cite{gpipe,megatron,zero,megatron2,megascale} specifically designed for the supervised training paradigms, such as pretraining and supervised fine-tuning~\cite{raft, zhang2024survey}.
% Meanwhile, the critical fine-tuning technique, known as Reinforcement Learning from Human Feedback (RLHF), catalyzed the evolution of GPT-3 into ChatGPT~\cite{llm-rlhf-openai-1,llm-rlhf-openai-2,instrgpt}. 
Meanwhile, another remarkable training paradigm for LLMs, known as Reinforcement Learning from Human Feedback (RLHF), is the foundation technique for the success of ChatGPT-like models~\cite{llm-rlhf-openai-1,llm-rlhf-openai-2,instrgpt,gemini,claude,llama2,harmlessllm-rlhf,openaio1}.
The workflow of RLHF training is much more complicated than supervised training. 
However, most existing RLHF systems adopt parallelization techniques directly from supervised training~\cite{dschat,openrlhf,nemo-aligner}, which could lead to sub-optimal training efficiency.
% Despite RLHF's crucial role in production-level LLM applications~\cite{claude,llama2,harmlessllm-rlhf,gemini,openaio1}, research regarding developing an efficient RLHF system is largely missing.

% The workflow of RLHF training is {much more complicated than supervised training for LLMs}. %as follows. 
The typical workflow of RLHF, which is often based on Proximal Policy Optimization (PPO)~\cite{ppo} algorithm, involves three distinct types of computational tasks on four LLMs with independent parameters.
In each RLHF training iteration, a primary LLM (the training target, referred to as the \emph{Actor} model) receives prompts and generates responses {(i.e., the generation tasks)}. These responses are evaluated by three additional LLMs: the \emph{Reward} model, the \emph{Reference} model, and the \emph{Critic} model {(i.e., the inference tasks)}. Then, \emph{Actor} and \emph{Critic} use the evaluation results to compute gradients and update their parameters (i.e., the training tasks).

%, namely \yw{iteratively} computing loss, calculating gradients with back-propagation, and updating parameters). 
% In summary, {the workflow of RLHF contains} %There are 
% four LLMs with independent parameters (referred to as \textbf{\textit{models}}) and distinct types of computational tasks on GPUs (referred to as \textbf{\textit{model function calls}}), namely \emph{generation}, \emph{inference}, and \emph{training}.

\begin{figure*}
    \centering
    \includegraphics[width=0.85\textwidth]{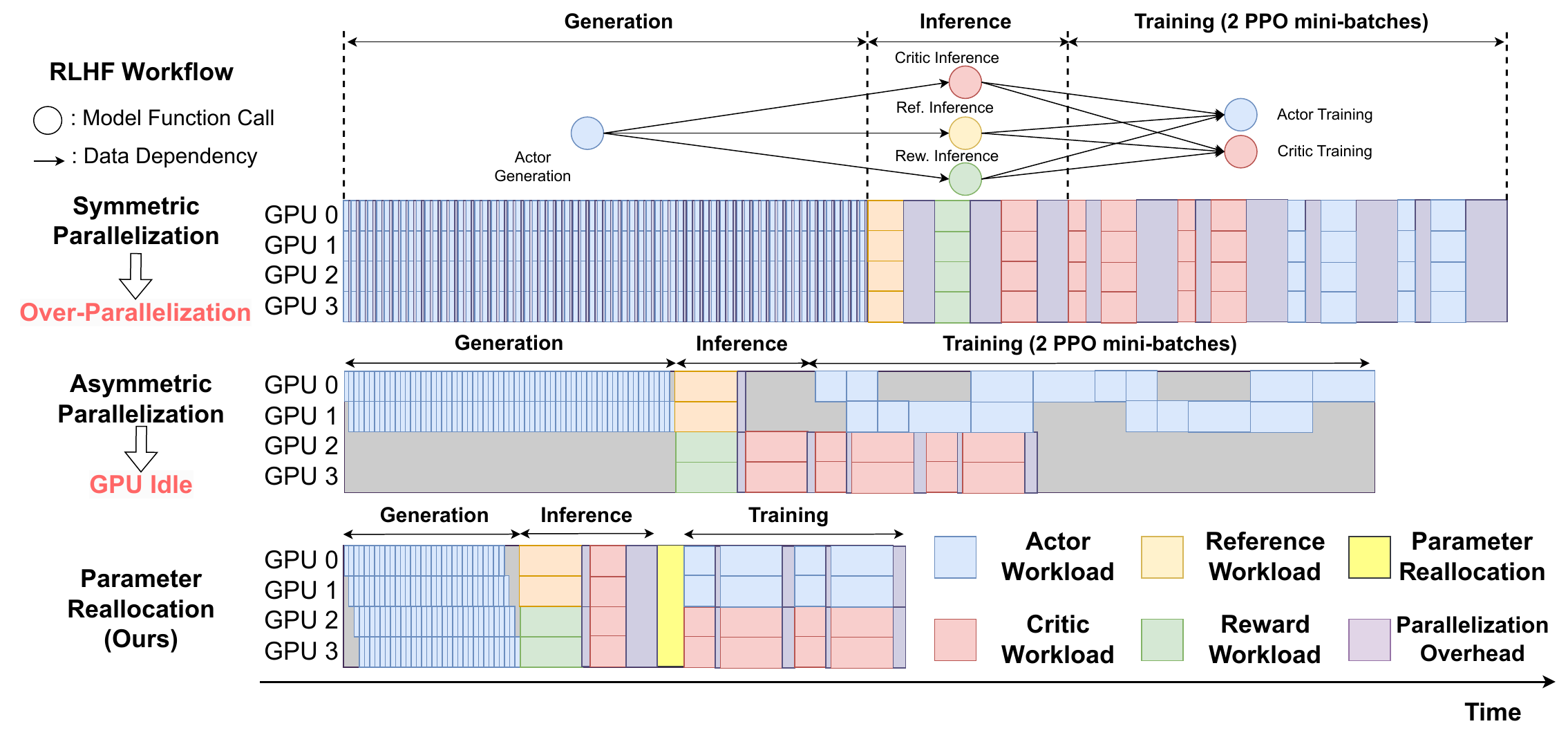}
    \vspace{-4mm}
    % \missingfigure{DeepSpeed execution instance vs ours}
    \caption{\small{An RLHF iteration breakdown based on the profiling of real systems (\Cref{tab:wall-time-breakdown}).
    The directed acyclic graph shows the RLHF workload.
    Nodes represent model function calls and edges represents their data dependencies. 
    We present timelines to visualize execution plans that employ: [top] the same parallelization strategy that spreads across the entire cluster for all LLMs, [middle] independent resource allocations and parallelization strategies for each LLM, and [bottom] distinct {resource allocations and} parallelization strategies for each \emph{model function call} generated by \sysname. The plan of \sysname considers parameter reallocation for the actor and critic model.}
    }
    \vspace{-4mm}
    \label{fig:intro}
\end{figure*}

To identify the drawbacks of the existing RLHF systems, we conduct a thorough profile and discover two major limitations.
% First, when the system adopts a \emph{symmetric parallelization} strategy (i.e., models are distributed to every GPU node that applies the same parallelization strategy), it is often \emph{over-parallelized}. 
{First, we note that many systems apply the same parallelization strategy that spreads across the entire GPU cluster for all LLMs.
\iffalse
\hz{``occupy'' may not be an accurate word here}
\fi
We name this a \emph{symmetric parallelization} strategy, which often leads to \emph{over-parallelization}.}
Our system profiling in~\Cref{fig:intro} (top) shows that over-parallelization leads to substantial synchronization and communication overheads (the light purple bars), thus compromising the end-to-end system performance. Moreover, different computational tasks are better off with different parallelization strategies~\cite{puzzle-rlhf}. A single global parallelization strategy, therefore, is likely to be sub-optimal. 
Accordingly, some other systems choose to allocate different LLMs to different sets of GPUs with different parallelization strategies.
In this way, tasks from different LLMs could be executed concurrently.
% An alternative way is to assign different models to different GPU nodes, where models can execute concurrently and apply different parallelization strategies independently. 
We call this an \emph{asymmetric parallelization} strategy.
However, our second observation is that such a strategy often causes under-utilization of the GPUs (e.g., the gray areas in ~\Cref{fig:intro} (middle)) because of the dependencies between tasks.

% While existing RLHF systems directly adopt parallelization techniques from supervised training~\cite{openrlhf,dschat}, it can usually lead to inferior performance, mainly due to the variety of computation tasks and intricate dependencies among multiple LLMs.
% Here, we emphasizes two key observations obtained by profiling previous RLHF systems.
% Our first observation is that \emph{symmetric parallelization} inflicts substantial communication overhead, which we call \emph{over-parallelization} (\Cref{fig:intro} top).
% Symmetric parallelization means distributing all models across the entire cluster with the same parallel strategy. 
% However, the optimal parallel strategies for different computation tasks can be different, as shown in \Cref{table:function_call_time}.
% A single global parallel strategy could exhibit overly large parallel degrees or create unnecessary synchronizations, both of which significantly contribute to communication overhead. 
% Our second observation is that \emph{asymmetric parallelization} causes GPU idle (\Cref{fig:intro} middle).
% Asymmetric parallelization individually assigns and parallelizes models, allowing concurrent execution of model function calls on a subset of devices.
% In this situation, when dependencies are not met, GPUs are forced to wait, resulting in low GPU utilization.

\begin{figure}
    \centering
    \includegraphics[width=0.87\columnwidth]{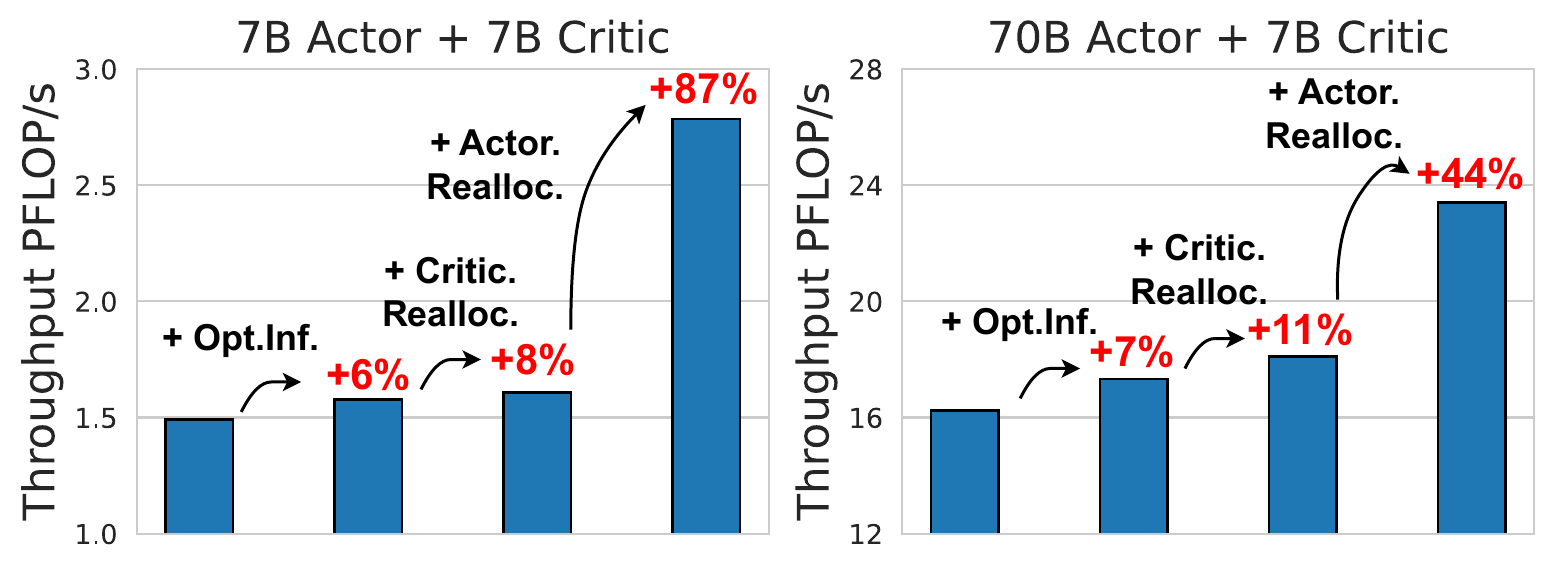}
    \vspace{-4mm}
    \caption{\small{The optimization opportunity over a 3D parallelism execution plan inspired by pre-training. We show the sequential improvement by optimizing inference parallelization strategy, reallocating critic workloads (inference/training), and reallocating actor workloads (generation/training).}}
    \vspace{-8mm}
    \label{fig:intro-optimization-space}
\end{figure}

% The crux of the above inefficiencies is that {the allocation of LLMs on GPUs is fixed throughout training}, which implies a fixed parallelization strategy as well. 
The crux of the above inefficiencies is that resource allocations and parallelization strategies for LLMs are fixed throughout training.
Therefore, we propose to enable \emph{dynamic reallocation of model parameters} between GPUs, allowing fine-grained resource allocations and parallel strategies at the task level to improve the efficiency of the entire RLHF training process.
For clarity, we refer to an individual task on an LLM as a \emph{model function call}.
As shown in ~\Cref{fig:intro} (bottom), by {first} choosing a parallelization strategy tailored for each model function call (e.g., Actor generation and training) and then executing these calls concurrently with a smaller parallelization degree (e.g., Actor and Critic training),
\iffalse
\hz{I don't know where to look in ~\Cref{fig:intro} (bottom) for the ``tailored strategy'' and ``a smaller parallelization degree''. Describe what actually happened there if space permits?}
\fi
we can reduce the communication overhead caused by parallelization while maximizing GPU utilization.
Parameter reallocation effectively addresses the limitations of prior solutions and can lead to a significant end-to-end throughput improvement, as we show in \Cref{fig:intro-optimization-space}.

Based on the key idea of parameter reallocation, we developed \sysname, a pioneering system for efficient RLHF training. 
\sysname consists of two components, i.e., an execution plan generator and a runtime engine.  
% An execution plan is formulated as an augmented dataflow graph, which specifies a particular execution strategy of the RLHF training workflow given the desired algorithmic and hardware configurations. 
An execution plan specifies the resource allocations and parallelization strategies for every model function call in the RLHF training workflow under a specific algorithmic and hardware configuration.
The execution plan generator performs Markov Chain Monte Carlo (MCMC) sampling to search for efficient execution plans using an extremely lightweight profiling-assisted runtime estimator.
After a sufficiently good execution plan is obtained, the runtime engine deploys the derived plan by effective parallelization and parameter redistribution. 

Our experimental evaluation entails RLHF training on LLaMA models~\cite{llama2,llama3} ranging from 7 to 70 billion parameters across 8 to 128 Nvidia H100 GPUs. 
Results showcase that \sysname is able to achieve a speedup up to 3.58 times over the baseline systems. 
Furthermore, we demonstrate that the performance of \sysname's searched execution plans surpasses heuristic plans based on Megatron by 54\% on average and up to 81\% with a longer context length.

%We speedup the searching process by carefully designing solution spaces and applying a light-weight cost estimator to evaluate the time and memory cost of candidate execution plans
% To speedup the searching process, we use several carefully designed constraint to shrink the solution space without hurting the performance of searched plan, and a light-weight cost estimator to evaluate the time and memory cost of candidate execution plans. 

% In the formulation, we consider RLHF training as a dataflow graph, and its execution plan an augmented dataflow graph.

In summary, our contributions are as follows:
\begin{itemize}[leftmargin=5mm, topsep=0mm]
\setlength\itemsep{0pt}
\item {We propose to reallocate model parameters dynamically for efficient RLHF training.}
%We introduce a novel problem formulation to seek a fast execution plan for the RLHF workflow, additionally considering parameter reallocation that potentially improves system throughput.
\item {We introduce execution plans at the model function call level and propose an efficient search algorithm to identify fast plans.}
% \hz{Is ``execution plan'' really a concept you invented? I would just go with ``We introduce execution plans at the model function call level and propose an efficient search algorithm to identify fast plans.''}
\item We design and implement \sysname, an RLHF training system that can automatically discover and run a fast execution plan with a high {training} throughput.
\item 
We conduct comprehensive evaluations with detailed breakdowns and ablation studies. \sysname achieves up to $3.58\times$ higher throughput compared to the baselines. % that show the superior performance ($2$-$10.6\times$) of \sysname compared to baseline systems.
%Furthermore, we present detailed breakdowns, ablation studies, and case analyses to elucidate \sysname's performance enhancements.
\end{itemize}

\section{Background}

\subsection{Introduction to RLHF}
\label{sec:background_rlhf}

% \outline{Introduction to RLHF algorithm on LLM. Introduce RLHF workflow (in a system view) in details. Maybe include brief introduction to other training methods such as SFT and DPO.}

% At a high level, RLHF aims to train a Large Language Model (LLM) to generate responses that score highly according to human preference, reflected by the Reward model.
{For the ease of illustration,}
this section adheres to the common practice of RLHF, focusing on GPT-like LLMs~\cite{gpt2,gpt3} and the Proximal Policy Optimization (PPO) algorithm~\cite{ppo}.
{However, we remark that \sysname can also support other RLHF algorithms, as we will discuss in \Cref{sec:solution_space}.}

An RLHF training iteration involves six model function calls on four LLMs: Actor generation, Reward inference, Critic inference, Reference inference, Actor training, and Critic training.
Their dependencies are shown in \Cref{fig:intro} (top).
% We further introduce these three types of function calls in details.
% A complete RLHF training iteration can be described as follows.
% Initially, a batch of prompts is sampled from the dataset, each composed of multiple tokens or integers. Subsequently, the Actor generates a response for each prompt.
In these model function calls, \emph{Generation} is composed of multiple forward passes. It involves a prefill phase and a decoding phase.
The prefill phase is a single forward pass, which consumes all prompt tokens to sample the first generated token. The decoding phase repeatedly inputs the (single) latest generated token and produces the subsequent token until termination.
\emph{Inference} is a forward pass over the combination of prompts and generated responses.
\emph{Training} is an ordinary supervised training iteration, composed of a forward pass, a backward pass, and a parameter update.
The next RLHF iteration then applies the updated Actor and Critic for generation and inference.
% Generated responses are evaluated by the Reference, Reward, and Critic models. These evaluations entail neural network \emph{inferences} or forward passes. Evaluation results are then aggregated and utilized as input for the \emph{training} of the Actor and Critic, which is similar to supervised training and composed of a forward pass, a backward pass, and parameter update.
% We have summarized this procedure as six \emph{model function calls} as shown in \Cref{fig:rlhf}.

% Subsequent iterations involve sampling new prompts, generating responses, and evaluting responses with the latest Actor and Critic.
% This entails partitioning training data into disjoint parts, with sequential forward-backward passes and parameter updates performed on each part.

Notably, training the Actor and Critic with PPO can incorporate multiple minibatches~\cite{instrgpt}.
For each minibatch, the parameter update must occur before the subsequent forward pass, distinguishing this approach from gradient accumulation that performs a single parameter update across minibatches.
% RLHF usually requires multiple consecutive training trials, and each trial is composed of multiple training iterations.
% For example, \citet{llama2} performed $4\times400$ RLHF iterations to build LLaMA-2 series.
%With the proprietary system in Meta, they report that a single RLHF iteration over the 70B model consumes about 330 seconds, resulting in a total of about 150 hours of training.
% They report that a single RLHF iteration over the 70B model in their proprietary system consumes about 330 seconds, resulting in about 150 hours of training in total.

\subsection{Parallelization of Large Language Models}

% \paragraph{Parallelization for LLMs.}
Classical parallelization approaches for LLMs encompass data, tensor-model, and pipeline-model parallelism.
% In these approaches, $N$ processes are launched on $N$ GPUs, with each process exclusively utilizing a single GPU.
% We first discuss them independently and then illustrate how to effectively combine them.
% \Cref{fig:parallel-intro} presents visualizations of these parallelization methods.

\emph{Data Parallelism (DP)} partitions data along the batch dimension and dispatches each partition to a model replicate
for independent computations.
After the backward pass during training, all DP peers should perform an all-reduce over gradients before applying them for parameter update.
% DP will consume a large amount of GPU memory due to duplicated parameter storage.

\emph{Tensor-model Parallelism (TP)} partitions model parameters and distributes matrix multiplications across multiple GPUs. Each TP rank processes the same data and produces a partial intermediate value.
Then, all TP peers perform an all-reduce over this value to obtain the full result and pass it to the next layer.
Since all TP peers should perform the all-reduce operation in each layer of the LLM, TP leads to substantial data communication overhead when scaling to more GPUs and deeper models.

\emph{Pipeline-model Parallelism (PP)} clusters adjacent layers into several \emph{pipeline stages}. PP peers transfer intermediate results among stages for a complete forward or backward pass, which entails less communication overhead than TP.
To improve the efficiency of PP, a common approach is to divide the data into micro-batches, allowing different GPUs to process different micro-batches simultaneously. 
\iffalse
\hz{Just to make sure: micro-batches and minibatches are different things?} \fuwei{true, we have only used "minibatch" for the PPO algorithm}
\fi

Since the above parallelization approaches are mutually independent, Megatron-LM~\cite{megatron2} integrates them as \emph{3D Paralleism} to perform LLM supervised training at scale. A \emph{parallelization strategy} $S$ is denoted by three integer values $(dp,tp,pp)$, representing the degrees of DP, TP, and PP, respectively.
Each coordinate in this grid represents a process running on an independent GPU.
3D parallelism entails near-optimal parallelization for GPT-like language models, which has been extensively experimented in previous studies~\cite{alpa}.

\section{Overview}
\label{sec:overview}

\begin{figure}[t!]
    \centering
    \includegraphics[width=0.87\columnwidth]{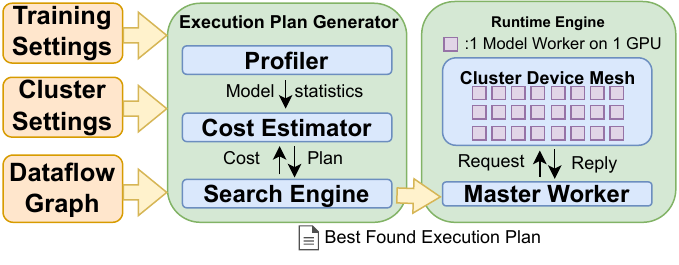}
    % \missingfigure{system architecture}
    \vspace{-3mm}
    \caption{\small{An overview of the architecture of \sysname.}}
    \vspace{-7mm}
    \label{fig:overview}
\end{figure}

% \todo{make the figure prettier}

% \todo{smaller size \cref{fig:overview}}
\sysname is a system capable of automatically planning and executing RLHF training workflows given algorithm and cluster specifications.
%The key idea behind is exploring the possibility of \emph{parameter reallocation}. 
%\mzy{It facilitates \sysname to produce an execution plan that assigns an independent resource allocation and parallelization strategy to \emph{each} model function call, which could largely reduces the communication overhead caused by parallelization and maximizes GPU utilization.}
\arxiv{
The key idea behind the design of \sysname is \emph{parameter reallocation}: dynamically reallocating model parameters across GPUs and assigning different GPU resources with a suitable parallelization strategy to each model function call.
Parameter reallocation enables \sysname to carry out the fine-grained orchestration of model function calls.
Specifically, \sysname can allocate distinct groups of resources to different model function calls, allowing them to execute concurrently on different sets of GPUs. \sysname also chooses a tailored parallelization strategy for each model function call. 
In this way, \sysname reduces communication overhead and improves GPU utilization without exceeding the memory limitation of the devices. 
By exploiting parameter reallocation, \sysname adopts a comprehensive design that addresses various training scenarios, seizing more optimization opportunities throughout the RLHF training workflow compared to existing systems~\cite{dschat, openrlhf, nemo-aligner}.
}

% Specifically, this execution plan assigns an independent resource allocation and parallelization strategy to \emph{each} model function call. 
% It then dynamically redistributes parameters at runtime to maximize the overall efficiency.

We summarize the steps of running \sysname as follows.
First, \sysname parses the RLHF workflow into a dataflow graph at the 
granularity of model function calls. 
%Then, a fast execution plan is produced by a specialized search algorithm to decide the parallel strategies of model function calls and intermediate data/parameter communications.
%Then, a specialized search algorithm produces a fast execution plan to decide the parallelization strategies and intermediate data/parameter communications.
Then, \sysname adopts an efficient search algorithm to produce a fast execution plan that includes parallelization strategies and intermediate data/parameter communications.
%Finally, \sysname runs this fast execution plan with an efficient implementation of the worker-based runtime engine.
Finally, \sysname runs this plan with an efficient worker-based runtime engine.

As demonstrated in \Cref{fig:overview}, there are two major components in the system, the \textit{\textbf{Execution Plan Generator}} and the \textit{\textbf{Runtime Engine}}.
The search engine in the execution plan generator continuously searches for execution plans with the Markov Chain Monte Carlo (MCMC) algorithm. 
A lightweight estimator calculates the approximate time cost of the proposed plan 
\iffalse
\hz{what's a ``searched plan''? Do you mean each candidate plan during the search?}
\fuwei{no, we compute cost for each proposed plan and accept or reject it according to the cost.}
\fi
by exploiting execution statistics of profiling. 
After reaching the search time limit, the fastest execution plan obtained is presented to the runtime engine for deployment.

The runtime engine is composed of a centralized master worker and multiple model workers.
The master worker resolves task dependencies and sends requests to the corresponding model workers for task execution.
% Each model worker is hosted on a single GPU, but it can hold multiple LLM handles (e.g., both Actor and Reward). 
Model workers act as RPC servers and respond to the master worker to update dependencies for subsequent requests.
% After completing the requested task, the model worker responds to the master worker to update dependencies for subsequent requests.
The interaction between the master worker and model workers repeats until the execution plan finishes.

\begin{figure}[t!]
    \centering
    \includegraphics[width=\columnwidth]{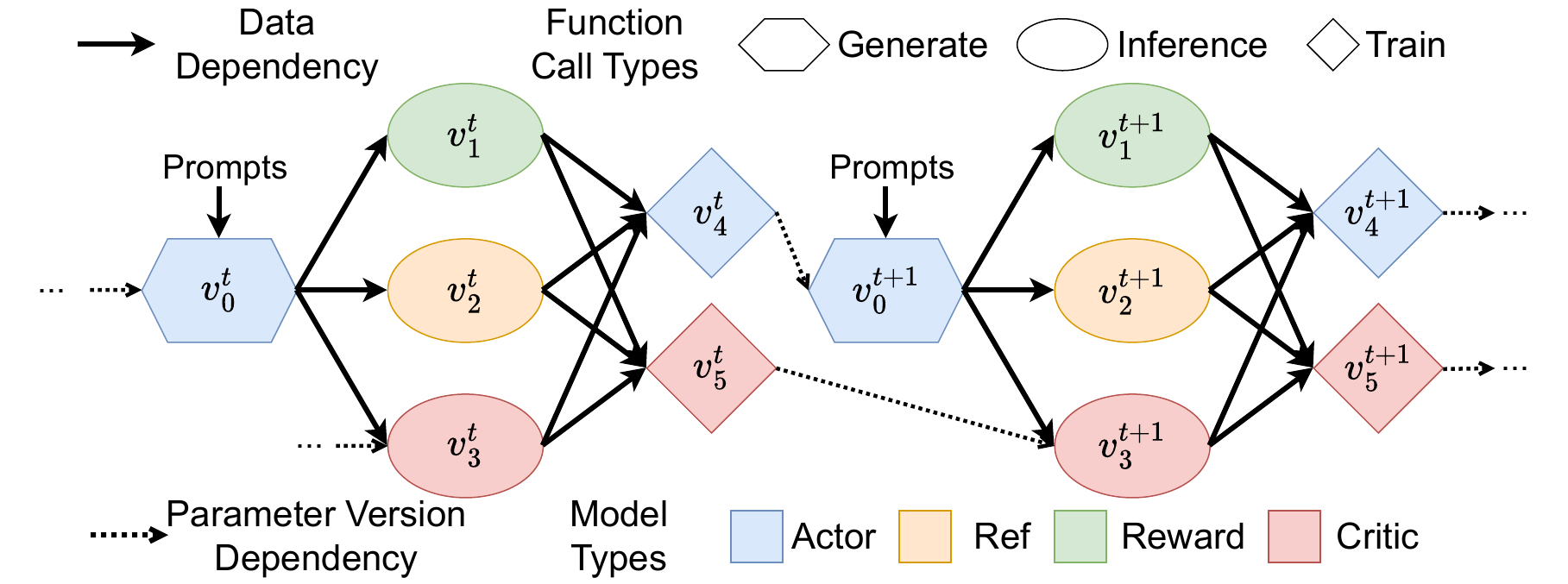}
    % \missingfigure{RLHF training iteration}
    \vspace{-8mm}
    \caption{\small{The dataflow graph of two consecutive RLHF iterations. Each \emph{\textbf{model}} is an independent LLM. Each \textbf{\emph{model function call}} is computational task of the model.}}
    \vspace{-8mm}
    \label{fig:rlhf}
\end{figure}

% \mzy{NOTE: moved API and code example to appendix}

\iffalse
We organize the rest of the paper as follows. 
\Cref{sec:solution_space} introduces our problem formulation and related definitions. 
\Cref{sec:method} explains the details of the execution plan generator. 
\Cref{sec:implementation} mainly introduces the implementation details of the runtime engine in \sysname. 
\Cref{sec:limit} discusses the advantages and limitations of \sysname.
\Cref{sec:experiments} shows our experiment results and ablation studies.
The final two sections discuss the related works and conclude the paper.
\hz{This paragraph is unnecessary}
\fi

% \outline{\sysname  is a training system that is capable of 1.automatically finding an optimized parallelization strategy and scheduling for large language model RLHF (and other multi-model algorithms) and 2. efficiently execute the training workflow according to the strategy. }
% \todo{architecture main figure}

\section{Problem Formulation}
\label{sec:solution_space}

\sysname is designed to accelerate RLHF workflows with GPT-like LLMs. To achieve this goal, we introduce the concept of execution plans at the model function call level. We formulate the problem as: taking training configurations (e.g., model size and batch size) and cluster specifications as inputs, search for an optimized execution plan that is able to be executed on the given distributed cluster.  
% The workflow is structured as a Directed Acyclic Graph (DAG). The nodes represent training, inference, or generation function calls. The edges denote data or parameter version dependencies. Inputs to \sysname include the DAG, training configurations (e.g., model size and batch size), and cluster specifications. \sysname outputs an optimized execution plan and executes it across a distributed cluster.
% \sysname aims to find a fast execution plan for RLHF, given the training configurations (e.g., size for each model and training batch size) and the cluster specifications.
In this section, we introduce our detailed {terminology definitions} in our formulation of the execution plan search problem. % and definitions of related concepts. 
% and the solution space explored by \sysname. 
% It is worth noticing that the following formulation is not only suitable for RLHF, but also capable of characterizing other training algorithms that incorporates multiple model function calls on different LLMs. 

% \subsection{Execution Plans of Dataflow Graph}
\textbf{{Dataflow Graph.}}
\sysname considers the workflow of RLHF {training} as a dataflow graph $\mathcal{G} = (\mathcal{V}, \mathcal{E})$, as demonstrated in \Cref{fig:rlhf}.
A node $v_{i}^{t}\in \mathcal{V}$ represents the $i$-th model function call at the $t$-th training iteration.
An edge $(v, v^\prime) \in \mathcal{E}$ indicates a data or parameter version dependency.
We emphasize that $\mathcal{G}$ represents the {concatenated graph} of all {the} iterations {throughout the entire training process}. %rather than a single one.
By operating on $\mathcal{G}$, we can potentially overlap computations with no mutual dependencies {across training iterations}. % end-to-end latency.
% In each training iteration, each type of model function call is executed once, and the data dependencies between them are identical across different iterations. 
% Dataflow graphs of consecutive training iterations are concatenated with dependencies caused by parameter updates.  

\begin{figure}[t!]
    \centering
    \includegraphics[width=\columnwidth]{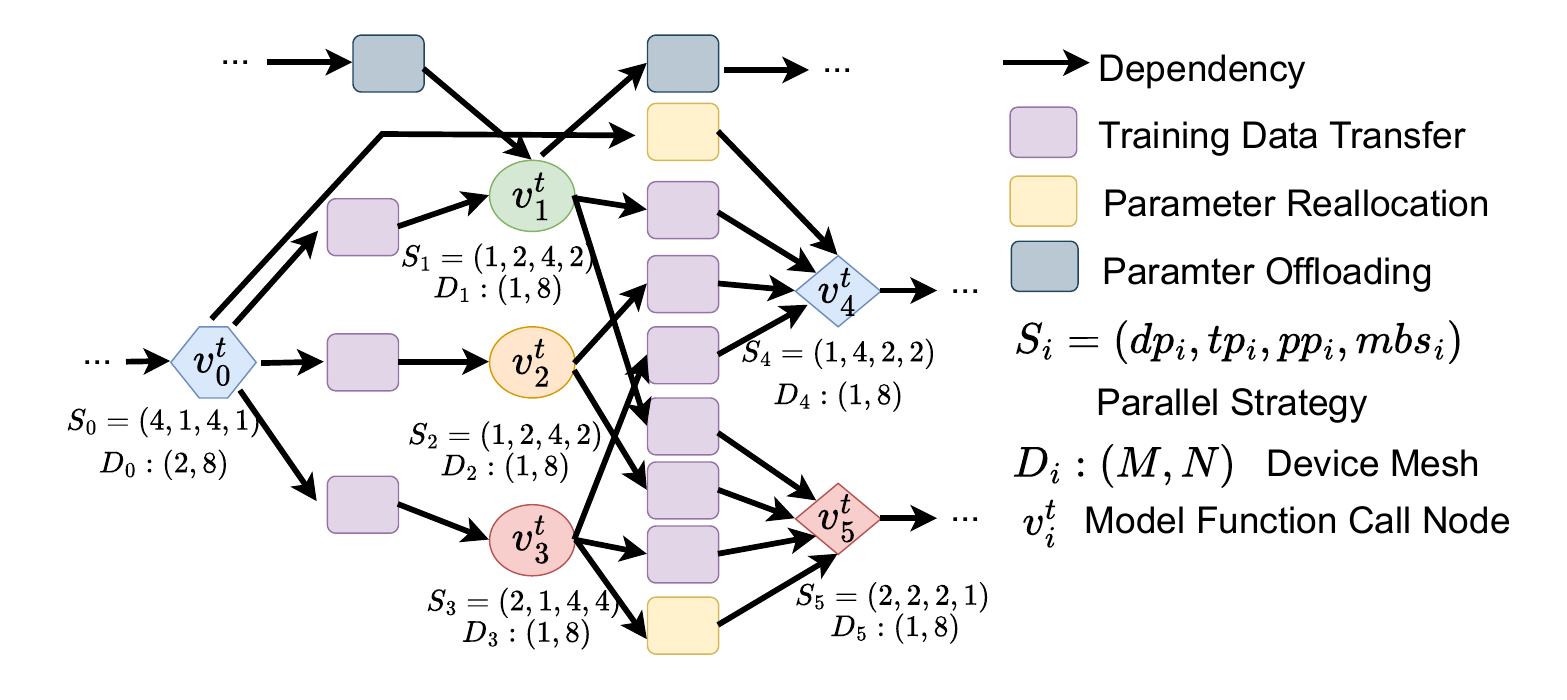}
    % \missingfigure{An execution plan instance}
    \vspace{-8mm}
    \caption{\small{An augmented dataflow graph $\mathcal{G}_p$ of an execution plan instance $p$ in the $t$-th RLHF iteration.}}
    \vspace{-7mm}
    \label{fig:execution_plan}
\end{figure}

% o characterize physical devices in the cluster, we follow the definition of device mesh in Alpa~\cite{alpa}. 
% , in which $\mathcal{D}$ denotes all feasible choices of device meshes in the cluster.
% Note that two device meshes with the same shape are not guaranteed to be on identical cluster hosts.
% Communication bandwidth between GPUs within a cluster nodeany rows or device within one row are assumed to be identical.

\textbf{{Device Mesh.}}
% A device mesh $D$ is the unit for executing an individual function call. It is defined as a two-dimensional grid of GPUs located in the cluster. 
A device mesh $D$ is defined as a two-dimensional grid of GPUs.
The shape of $D$ is denoted as $(N, M)$ if it covers $N$ nodes equipped with $M$ devices. Note that device meshes with the same shape could have different locations. 
We assume all devices in the cluster have the same computing capability with identical intra-node bandwidths, and inter-node bandwidths.
\iffalse
\hz{I don't think this is the correct way of using the word ``respectively''.}
\fi

% Different device meshes can overlap if they share some devices; otherwise, they are disjoint. We assume all devices in the cluster have the same computing capability. 

% It is worth mentioning that different from device mesh in \cite{alpa}, devices in a row shares a communication bandwidth to memory cache device located on the physical node, such as CPU memory and NVMe. 
% This enables \sysname to explore the possibilities of parameter offloading in our solution space.

\textbf{{Execution Plan.}}
An execution plan $p$ of a dataflow graph $\mathcal{G}$ assigns a device mesh $D_{i}$ and parallelization strategy $S_{i}$ for the $i$-th \emph{individual function call} in $\mathcal{G}$. 
We express an execution plan $p$ in the form of an \textbf{\textit{augmented dataflow graph}} $\mathcal{G}_p=(\mathcal{V}_p, \mathcal{E}_p)$, as visualized in \Cref{fig:execution_plan}.
It also involves data transfer, parameter redistribution, and offloading~\cite{zero-offload} across function calls.
They are represented as a set of extra nodes in $\mathcal{G}_p$ (rounded squares in \Cref{fig:execution_plan}).

\textbf{{Search Space.}}
We make several assumptions to make the generation and deployment of execution plans practically feasible.
% To limit the possible choices of the execution plan for an experiment in \sysname, we apply several constraints to $D_{i}$ and $S_{i}$ while maintaining the capability of \sysname to obtain an optimized solution. 
% The search space of the execution plans is carefully designed to limit the number of possible choices to shorten the search time, and ensure the capability of the search engine to produce an optimized solution under various circumstances. 
First, we assume that $D_i$ either covers several entire hosts or a consecutive portion that is capable of dividing the number of devices on one host, e.g., (1, 1), (1, 2), (1, 4), (1, 8), (2, 8), $\cdots$, ($N$, 8) in a cluster of $(N, 8)$. 
This ensures that multiple device meshes can fully cover the entire cluster, eliminating sub-optimal execution plans with idle GPUs~\cite{alpa}.
Second, $S_i$ considers the 3D parallelism degrees $(dp_i,tp_i,pp_i)$ and the number of micro batches $mbs_i$.
Data will be divided into $mbs_i$ portions and passed to the function call sequentially.
This feature provides an option to avoid the out-of-memory issue with a large batch size and context length. 

\textbf{{Beyond PPO.}}
{The example shown in \Cref{fig:rlhf} and \Cref{fig:execution_plan} represents the typical RLHF algorithm, PPO. Meanwhile, we emphasize that our formulation is inherently expressive for training algorithms whose workflow could be decomposed into the function calls and represented as a dataflow graph. Experiments in \Cref{sec:exp-beyond-ppo} demonstrate the capability of \sysname to accelerate other prevalent RLHF algorithms including DPO~\cite{dpo}, ReMax~\cite{remax} and GRPO~\cite{deepseekmath-grpo}.}

%the execution plans in \sysname are also capable of considering parameter offloading in runtime, which is subsumed into a special case of parameter reallocation with source or destination is a memory storage device attached to cluster nodes, such as CPU memory or NVMe. 

% We do not employ complex parallelization strategies at the granularity of operation is that it is unnecessary for GPT-like language models, which is the major focus of \sysname. 

%In \sysname, we only consider 3-D parallelization strategies, first introduced by Megatron~\cite{megatron}, of model function calls.   The reason why we do not employ complex parallelization strategies based on operation graphs is that it is unnecessary for GPT-like large language models, which is the major focus of \sysname.
\iffalse
\hz{This section is called ``Problem Formulation'', but I didn't find any problem statement (i.e., what is the problem you are trying to design an algorithm for?). The current section reads more like problem settings.}
\fuwei{the first paragraph has been rewritten}
\fi

\section{Execution Plan Generator}
\label{sec:method}

The execution plan generator takes the dataflow graph, the training configurations, and the cluster specifications as inputs to automatically search for a rapid execution plan in the form of an augmented dataflow graph.
This generator comprises two primary components.
First, a lightweight runtime estimator predicts the time and memory cost of any execution plan, leveraging statistical results from profiling.
Second, a search engine refines the proposed execution plan using a Markov Chain Monte Carlo (MCMC) search algorithm based on the preceding cost estimation.

\subsection{Estimation}
The architecture of LLMs is typically a stack of identical layers, exhibiting {clear computation patterns}. %high regularity. 
Hence, we can profile the time cost of operations on individual layers and estimate the total cost of each model function call through arithmetic operations. We present a lightweight runtime estimator assisted by profiling. Profiling the statistics in a single experiment takes only minutes, while evaluating the cost for a candidate execution plan requires only hundreds of microseconds, as opposed to several minutes for profiling a single plan on a real run. In the subsequent paragraphs, we denote the estimated values of the time cost and the runtime memory of an execution plan as $\textit{TimeCost}(\mathcal{G}_p)$ and $\textit{MaxMem}(\mathcal{G}_p)$.

\textbf{{Time Cost.}}
We first estimate the time cost for each node $v \in \mathcal{V}_p$.
For model function call nodes, \sysname profiles the cost of forward, backward, and associated communication (e.g., all-reduce) of individual layers across a set of data input sizes. 
The range of this set is decided by the configured batch size, the number of devices in the cluster, and the minimum batch size on each device according to parallelization strategies. 
We only profile sizes that are powers of two in this range. If the data input size for $v$ falls outside the profiling set, \sysname estimates the time cost using a linear interpolation of the existing profiling statistics. 
We estimate the costs of data and parameter transfer by running a simulation to the algorithm outlined in \Cref{sec:implementation}.
We approximate the time with the data size and the bandwidth instead of running a real NCCL operation.

Next, we derive $\textit{TimeCost}(\mathcal{G}_p)$ from the cost of each node. 
The calculation can be much more complex than simple summation because different nodes can be executed concurrently on disjoint device meshes.
We employ an algorithm to find the shortest path from source nodes to sink nodes in $\mathcal{G}_p$, with the constraint that nodes assigned to overlapped device meshes cannot execute simultaneously. The algorithm assigns each node $v \in \mathcal{G}_p$ with attributes \textit{StartTime}, \textit{EndTime}, and \textit{ReadyTime}. Each device mesh $D$ tracks the last completed node from all devices within $D$ as \textit{$D$.last}.
The algorithm maintains a priority queue containing all nodes {that have been ready for execution but not yet completed}. The priority queue {iteratively} selects the node with the minimum ready time, marks it as completed, updates \textit{$D$.last} for all $D$, and adds new ready nodes to the queue. When the priority queue {becomes empty}, all nodes in $\mathcal{G}_p$ should be completed, and the maximal \textit{EndTime} of all nodes yields the final result of $\textit{TimeCost}(\mathcal{G}_p)$. {The details of the simulation algorithm is shown by \Cref{algo:simulate} in \Cref{app:simulation_algo}.}

\textbf{{Maximum Memory Allocated.}}
An execution plan $p$ is executable only if its maximum runtime memory does not exceed device limitations. 
\iffalse
Memory allocation in \sysname follows these principles:
\begin{enumerate}[leftmargin=*,labelsep=0mm]
\item For a model designated for training, the gradients and optimizer states cannot be redistributed to other devices alongside the parameters.
\item Model parameters can be redistributed to the CPU memory or a different device mesh, freeing the memory occupied in their original location.
\item Memory for intermediate results, including KV cache, logits in model outputs, and intermediate activations, is dynamically allocated during execution.
\item The buffer memory required for data transfer is negligible compared to other memory costs.
\end{enumerate}
\fi
% To estimate maximum memory allocated across all devices when running an execution plan, 
We categorize the runtime memory into the {\textit{static memory}} and the {\textit{active memory}}. 
{The static memory consists of the gradients and optimizer states, which will not be freed or transferred until the entire experiment finishes. 
The active memory is only stored in GPU when it is required, including the KV cache, intermediate activations, and reallocable parameters, etc.
We first calculate the static memory and the peak active memory allocated for each function call according to their parallelization strategies. 
Afterwards, we calculate the peak memory during an RLHF iteration for each device and take the maximum to obtain $\textit{MaxMem}(\mathcal{G}_p)$.
}

\subsection{Execution Plan Search}

An execution plan $p$ assigns a device mesh $D_i$ and a parallelization strategy $S_i$ for the $i$-th model function call. The number of choices grows exponentially with the number of devices in the cluster. For instance, in a cluster of shape $(8, 8)$, there are over $500$ options for each model function call, and over $10^{16}$ execution plans in total, rendering brute-force enumeration practically infeasible. Therefore, \sysname employs an efficient MCMC-based search algorithm tailored for this problem setting.

We associate each execution plan with a cost defined by
\begin{footnotesize}
\begin{align*}
&cost(\mathcal{G}_p)= I\left(MaxMem(\mathcal{G}p) < mem_{d}\right) \cdot TimeCost(\mathcal{G}_p) \\
& + \left(1-I\left(MaxMem(\mathcal{G}p< mem_{d}\right)\right) \cdot \alpha
\cdot TimeCost(\mathcal{G}_p),
\end{align*}
\end{footnotesize}
where $mem_d$ is the device memory capacity, $I$ is an OOM indicator, and $\alpha$ is a large integer representing the OOM penalty. We then define an energy-based distribution $P(p) \propto \exp(-\beta \cdot cost(\mathcal{G}_p))$, where $\beta$ is the sampling temperature. Lower-cost execution plans have higher probabilities of being sampled from $P$. Hence, the {searching} process for a fast execution plan {becomes} drawing samples from the target distribution $P$, where MCMC techniques come into play.

We employ the Metropolis-Hastings algorithm~\cite{mcmc_metropolis, mcmc_hastings} for drawing samples from $P$. The sampling process begins with a greedy solution $p_0$ minimizing the summation of time costs of all function calls. Notably, this execution plan can be sub-optimal due to the excessive memory allocation on devices and the lack of overlap between different model function calls. Subsequently, we construct a Markov Chain comprising execution plans $p_0, p_1, \cdots$. We alter $D_i$ and $S_i$ of a random function call $i$ and accept this transition with probability
\[
P_{acc}(p_n\to p_{n+1})=\min\left(1,\frac{P(p_{n+1})}{P(p_{n})}\right),
\]
This process repeats until a terminating condition, such as when a constant time limitation is met. Finally, the execution plan with the minimum $TimeCost(\mathcal{G}_p)$ throughout the entire searching process is selected as the output of the execution plan generator.

\iffalse
We select a common approach \todo{cite} to convert the cost function into following target distribution, with a predetermined constant $\beta$:
\[P(p) \propto \exp({-\beta\cdot cost(\mathcal{G}_p)})\]
%I(\MaxMemo) \dot\text{TimeCost}(\mathcal{G}_p
We use the Metropolis-Hastings algorithm~\cite{mcmc_metropolis, mcmc_hastings} to create the Markov Chain.
The algorithm maintains a current execution plan $p$ and proposes a new one $p^{*}$ from a proposal distribution $g(p|p^{*})$. 
Then the new plan is accepted with a probability of $A(p, p^{*})$ which could be calculated by:
\[A(p, p^{*}) = \min(1, \frac{P(p^{*})Q(p|p^{*})}{P(p)Q(p^{*}|p)})\]
If $p^{*}$ fails to be accepted, another new plan is proposed.
We limit the difference between $p^{*}$ and $p$ to the device mesh allocation and parallelization strategy of a single model function call. 
Typically, the proposal distribution $g(p|p^{*})$ follows a uniform distribution across potential $p^{*}$. 
Nevertheless, alternative distributions could also be leveraged to optimize the search efficiency.
\fi

% Intuitively, the MCMC sampling greedily visits samples with frequencies proportional to their probability in the distribution.
% In other words, our search algorithm tends to move towards execution plans that have a lower cost, but could also escape local minima and explore 

\section{Runtime Engine}
\label{sec:implementation}

\begin{figure*}[t!]
    \centering
    \includegraphics[width=0.83\textwidth]{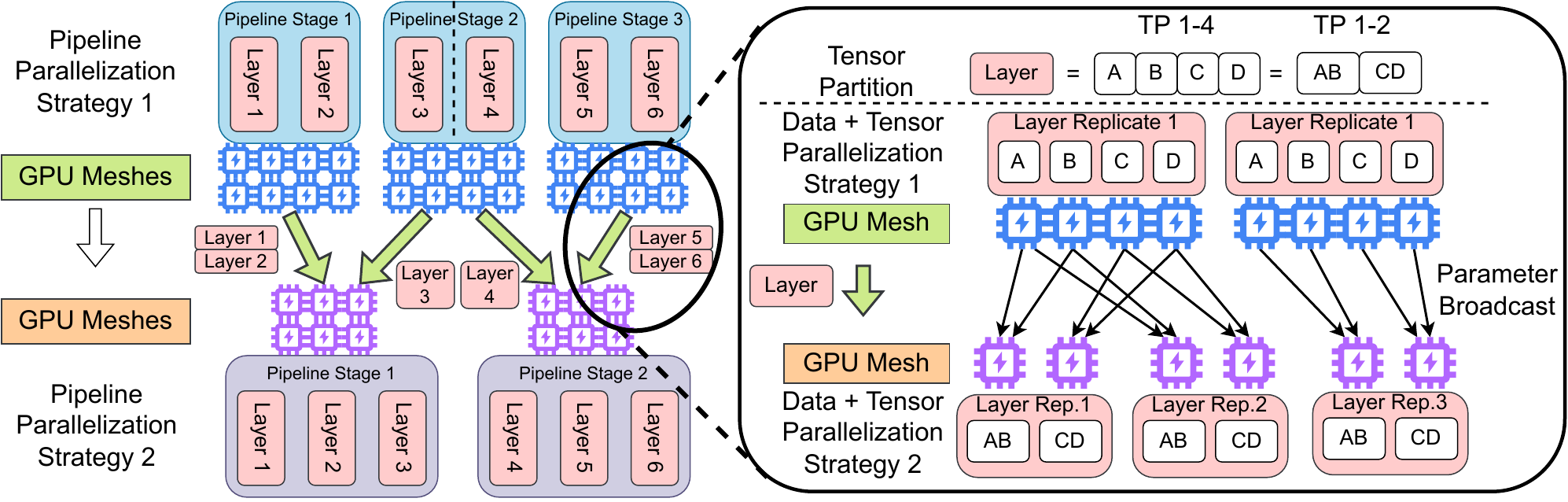}
    \vspace{-2mm}
    \caption{\small{The parameter redistribution is a hierarchical procedure. In the outer loop (left), each pair of pipeline stages communicates the parameters of their common layers. These parameters are distributedly stored in a DP plus TP device mesh. In the inner loop (right), layers are remapped from one DP plus TP mesh to another. Each destination GPU is assigned with a source that has the lowest communication cost. All assigned sources broadcast TP partitions required by destination GPUs in parallel.}}
    \label{fig:realloc}
    \vspace{-5mm}
\end{figure*}

In this section, we introduce the runtime engine, including the implementation details of workers, redistributing parameters, and transferring data among function calls.

\textbf{{Workers.}}
The master worker resides on a CPU and executes several \texttt{asyncio} coroutines to manage the each function call. The coroutine awaits the completion of all the parent function calls and dispatches requests via sockets upon the function call is ready. These messages do not transfer the associated data. Instead, the data is retained locally in the GPUs of model workers. The master worker communicates the data locations to the model workers in requests to initiate data transfers. Each model worker acts as an RPC server on a GPU. It polls requests from the socket for each local LLM handle (e.g., Actor and Reward) in a round-robin manner. Received requests are put in a FIFO queue for sequential execution and responding.
% \iffalse
% \paragraph{3D Parallelism}
% We employed the ZeRO DP implementation from DeepSpeed~\cite{deepspeed} to ensure a fair comparison with baselines.
% We follow the TP and PP implementation in Megatron-LM~\cite{megatron} and further implement pipeline schedules for generation and inference.
% \fi

% The runtime engine is majorly composed of two types of workers, the centralized master worker and model workers. 
% The model workers can perform model function calls in arbitrary 3D parallelization strategies. 
% We employ the ZeRO DP implementation from DeepSpeed~\cite{deepspeed} to ensure a fair comparison with baselines.
% As for TP and PP, we follow implementation in Megatron-LM~\cite{megatron} and further implement pipeline schedules for generation and inference.

\textbf{{Redistributing Parameters}} encompasses host-device (e.g., offload) and device-device communications. Host-device communication utilizes an additional CUDA stream for asynchronous memory copying. Device-device communication involves mapping one 3D parallelization strategy to another, e.g., from $(dp_1,tp_1,pp_1)$ to $(dp_2,tp_2,pp_2)$.
We regard the remapping as a hierarchical process consisting of an outer loop (\Cref{fig:realloc} left) and an inner loop (\Cref{fig:realloc} right). Initially, we focus on remapping pipeline stages from $pp_1$ to $pp_2$. Each stage $i\in[pp_1]$ holds a group of layers distributed in a device mesh specified by $(dp_1,tp_1)$. For each stage pair $(i,j)$, where $i\in[pp_1]$ and $j\in[pp_2]$, we transfer the parameters of common layers between device meshes specified by $(dp_1,tp_1)$ and $(dp_2,tp_2)$. We denote the devices in $(dp_1,tp_1)$ as source GPUs and $(dp_2,tp_2)$ as destination GPUs. For each destination GPU, we greedily assign a source GPU with the lowest communication cost (e.g., a local GPU has a lower cost than remote GPUs). Once assigned, the source GPUs broadcast parameters to the destinations in parallel. This process iterates until all stage pairs $(i,j)$ are covered.

% To facilitate both, we remap parameters into a contiguous memory block.

% We remark that device meshes used by these two parallelization strategies can be overlapped. In this case, assigning overlapped GPUs incurs the lowest communication cost, which simplifies to local memory copy.

{\textbf{Data Transfer Among Function Calls.}}
Model function calls produce disjoint data {partitions} along the {DP dimension}, while {replicating} the data along the {TP dimension}. This mirrors the communication pattern of redistributing parameters in the right part of \Cref{fig:realloc}, but with reversed TP-DP dimensions. Therefore, we employ the same broadcast-based algorithm for data transfer. 
% Given that data occupies far less GPU memory than parameters, we additionally maintain a local cache to store the received data, reducing redundant communication.

\textbf{Remark:}
\citet{reshard} explored a similar problem to data transfer in \sysname. 
% They further analyzed the efficiency of a broadcast-based approach over send-receive and gather-scatter alternatives, validating the rationality of our implementation.
In our paper, we do not focus on developing an optimal communication algorithm in such scenarios, as long as the cost is minor compared to other workloads in RLHF, as we will show in \Cref{fig:breakdown}.

\section{Discussions}
\label{sec:limit}

This section discusses the advantages and limitations of \sysname and clarifies the contexts where \sysname can be applied.
\sysname is a system that is applicable on accelerating RLHF workflows composed of training, inference, or generation function calls with GPT-like LLMs. 
% Compared to existing specialized RLHF systems and libraries,
Apart from its superb performance, \sysname has following advantages ($\blacksquare$) and limitations ($\diamond$):
\begin{itemize}[leftmargin=*]
\setlength{\itemsep}{0pt}
% \item[$\blacksquare$] \sysname supports 3D parallelism and automatic execution for RLHF, which largely improves system throughput and eliminates human efforts in production. However, neither of them was supported in prior RLHF systems.
% is capable of automatically finding a fast execution plan in a short time, largely eliminating human efforts.
% \item[$\blacksquare$] \sysname explores a novel technique, parameter reallocation, in LLM training workflows, which can introduce a wide range of new optimization opportunities.
% used to manually compare and choose from a wide range of systematic options that could occur in setting an RLHF experiments.
\item[$\blacksquare$] \sysname's method is orthogonal to advanced techniques for accelerating individual function calls (e.g., Paged-attention~\cite{vllm}) or fusing different function calls (e.g., RLHFuse~\cite{rlhfuse}).
These techniques can be integrated for better performance. 
\item[$\blacksquare$] \sysname can generalize beyond the workflow for PPO. It can also significantly accelerate various other prevalent RLHF algorithms, such as DPO~\cite{dpo}.
% \item While \sysname is specialized for RLHF training, its methods is applicable on other training methods~\cite{dpo} that comprises training, inference and generation on multiple LLMs. With that being said, \sysname could serve as a motivation for the development of novel training algorithm for LLMs in the future. 
% \item[$\diamond$] \sysname does not consider parallelization strategies that goes beyond 3D parallelism, which could lead to inferior performance on deep learning models other than LLMs.
% \item \sysname assumes a static dataflow graph composed of model function calls. It does not automatically partition the computation graph into model function calls.
\item[$\diamond$] \sysname requires predictable function calls to ensure the validity of cost estimation. An unstable cluster or dynamic workflow (e.g., the generation length varies significantly during training) can violate this assumption.
% when the cluster environment is not stable, or output sequence length of generation greatly varies.
% \item \sysname's method could not handle training workflows whose computation could not be naturally divided into model function calls. 
\item[$\diamond$] The searching of \sysname does not guarantee optimality despite producing plans \arxiv{that are fast and efficient in practice.}
% \item The search algorithm of \sysname has additional hyper-parameters, i.e., OOM penalty $\alpha$ and sampling temperature $\beta$, and does not guarantee optimality.
% \item \sysname does not optimizes the throughput of individual model function calls with advanced techniques, e.g., using paged-attention~\cite{vllm} for generation, while most them are complementary to this paper.
\end{itemize}
% \sysname is applicable on workflows explicitly defined with generation, inference, and training over multiple GPT-like models.
% The same function call on a specific model can only occur once in each iteration.
% Although \sysname surpasses over prior RLHF systems by these advantages, it currently has a few limitations:
% Despite these limitations, we will demonstrate that \sysname largely improves the throughput over prior RLHF systems and heuristic execution plans in \Cref{sec:experiments}.

\begin{figure}
    \centering
    \includegraphics[width=0.9\columnwidth]{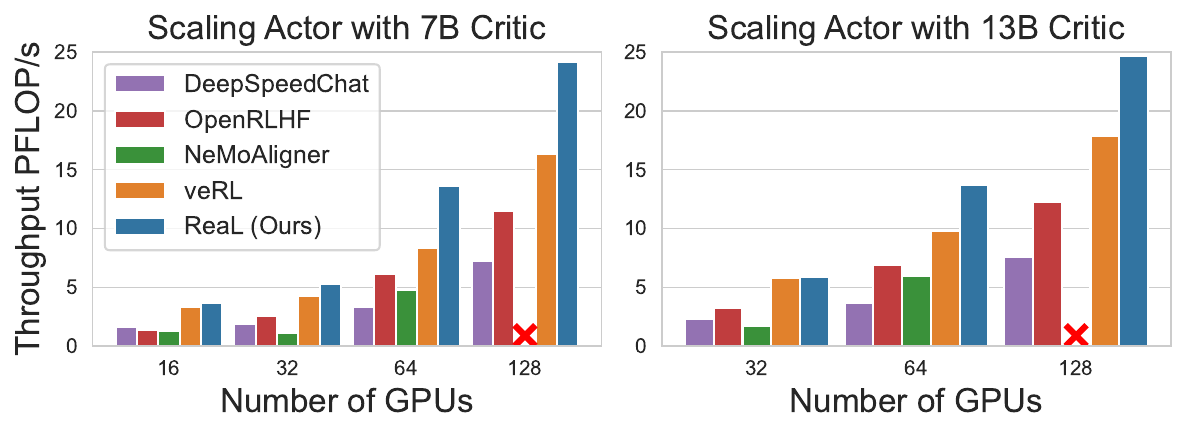}
    \vspace{-3mm}
    \caption{\arxiv{\small{An end-to-end throughput comparison with baseline systems. Red cross denotes instability and OOM errors caused by scalability issues.}}}
    \label{fig:other-sys-thpt}
    \vspace{-1mm}
\end{figure}

\section{Experiments}
\label{sec:experiments}

\sysname is implemented in Python (41k LoC) and C++ (2.5k LoC). The search engine and simulator components are written in C++, while the remaining modules leverage PyTorch~\cite{pytorch}.
Our experiments are conducted on a cluster of 128 H100 GPUs, interconnected via NVLink for the intra-node communication and RoCE with a 3.2Tbps bandwidth for the inter-node communication.
We adopt the most advanced LLaMA-3 models~\cite{llama3} for our experiments.
Since the vocabulary size of LLaMA-3 is large (128k), resulting a 250GB memory usage during computing softmax\footnote{$\textrm{VocabSize}\times\textrm{BatchSize}\times\textrm{CtxLen}\times\textrm{BytesPerParam}=128e3\times512\times2048\times2=250\textrm{GB}$}, we are only able to train a 70B actor with a 13B critic model under the resource constraint.

Our evaluation comprises five key components. First, we benchmark \sysname's end-to-end performance against two open-source RLHF systems and a heuristic baseline. Second, we present a detailed performance breakdown to identify key improvements. Third, we conduct an ablation study of the execution plan generator. Fourth, we demonstrate \sysname's compatibility with and acceleration of various RLHF algorithms beyond PPO. Finally, we analyze \sysname's strong scaling characteristics and provide suggestions for the practical usage.

%%%%%%%%%%%%%%%%%%%%%%%%%%%%%%%%%%%%%%%%%%%%%%%
\subsection{Comparison with Baselines}

\textbf{Baselines.}
We evaluate \sysname against four prominent open-source systems: DeepSpeed-Chat~\cite{dschat} (commit f73a6ed with DeepSpeed v0.15.1 as backend), \arxiv{veRL (Hybrid Flow)~\cite{hybridflow} (v0.2.0.post2 with vLLM v0.6.3 and FSDP on pytorch v2.4.0), NeMo-Aligner~\cite{nemo-aligner} (v0.4.0 with TRT-LLM v0.10.0 and Megatron v0.8.0)} and OpenRLHF~\cite{openrlhf} (v0.4.2 with vLLM v0.4.2 and DeepSpeed v0.15.0). 
Addtional details of the baseline systems are listed in \Cref{app:additional-baseline}.

DeepSpeed-Chat employs a symmetric parallelization strategy using the ZeRO-3 data parallelism~\cite{zero} across all RLHF models. Its \emph{Hybrid Engine} temporarily redistributes ZeRO-3 partitions to TP during the generation task, reverting afterward. Beyond this mechanism, DeepSpeed-Chat does not support TP or PP implementations. 

OpenRLHF implements an asymmetric parallelization strategy, dividing GPUs into three GPU groups. The groups hold the actor/reference model, the critic/reward model, and a generation engine using vLLM~\cite{vllm}.
The vLLM engine is only responsible for the actor generation. It remains idle during the actor training, awaiting parameter updates before proceeding to the next RLHF iteration. 

\arxiv{Similarly, NeMoAligner divides GPUs into two disjoint GPU groups. Unlike OpenRLHF, it locates actor training and generation on the same GPU group.
veRL (HybridFlow) is a concurrent work to \sysname that supports colocating models on GPUs and split placement of models on different GPU groups, including the strategies adopted by three previous systems.}

% While both DSChat and OpenRLHF represent specific instances of our execution plans, our search engine may not identify these configurations due to high estimated costs.
Additionally, we evaluate \sysname-Heuristic, a pre-training-inspired approach~\cite{megatron2} that implements a symmetric 3D parallelization across all models. This strategy combines the intra-node TP with the inter-node PP and DP, maximizing the DP degree within memory constraints.

\textbf{Settings.}
We evaluate the \emph{weak scaling} characteristics of \sysname, where the model size and the batch size 
%\footnote{\arxiv{OpenRLHF and DeepSpeedChat, limited to data parallelism, cannot efficiently operate with small batch sizes across multiple devices. We modify their code to support training in multiple micro batches to avoid OOM issues. veRL inherently supports this feature, but it still runs out of memory with large batch sizes, using the most conservative setting.}} 
both increase proportionally with the number of devices.
Our experimental configuration follows InstructGPT~\cite{instrgpt} with more details in \Cref{app:exp-detail}.

\textbf{Evaluation Metrics.}
Since the dataflow graph dependencies ensure consistent convergence properties, we focus on the total training throughput as our primary performance metric. Measurements are taken over 20 consecutive training iterations following appropriate warm-up periods. The observed throughput variation across trials is negligible, thus error bars are omitted from our figures.

\begin{figure}
    \centering
    \includegraphics[width=\columnwidth]{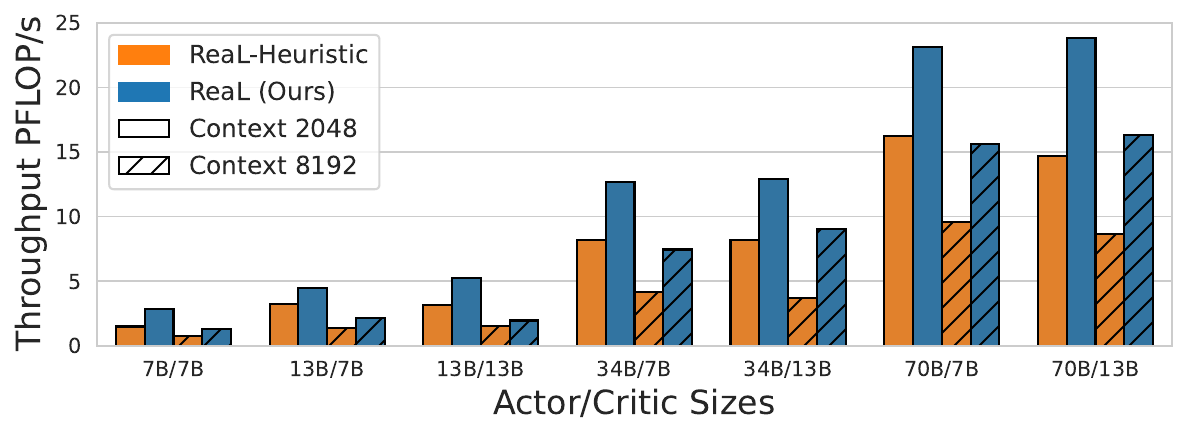}
    \vspace{-8mm}
    \caption{\small{Throughput comparisons with the heuristic execution plan with different context lengths.}}
    \vspace{-4mm}
    \label{fig:heuristic-ctx-bs}
\end{figure}

\textbf{Results.}
Throughput comparisons presented in \Cref{fig:other-sys-thpt} and \Cref{fig:heuristic-ctx-bs} demonstrate \sysname's superior performance. Compared to the baseline systems, \sysname achieves at most $3.58\times$ higher throughput. The search-generated execution plan outperforms \sysname-Heuristic by an average of $54\%$. This advantage further increases to $81\%$ when extending the context length from 2048 to 8192 tokens, highlighting \sysname's particular effectiveness in long-context scenarios.

\subsection{Breakdown Analysis}

\begin{figure}
    \centering
    \includegraphics[width=0.98\columnwidth]{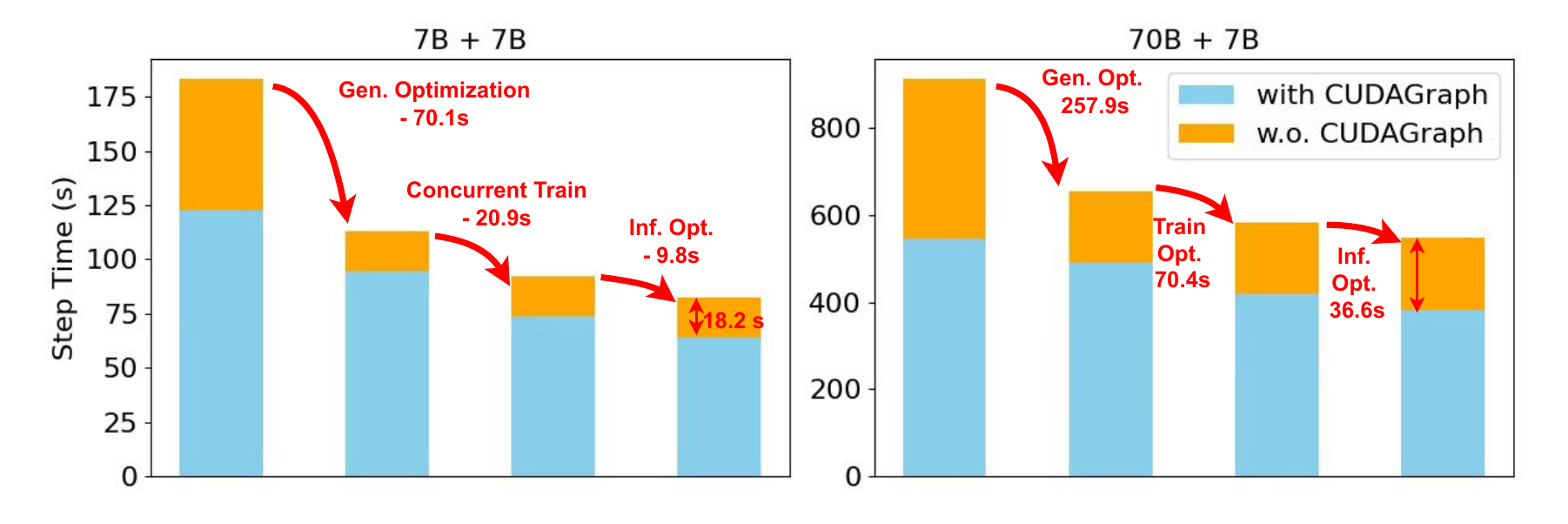}
    \vspace{-4mm}
    \caption{\arxiv{\small{The wall time of one training step of 7B + 7B and 70B + 7B settings, with different levels of optimization. 
    The left most and right most bars show the performance of \sysname-Heuristic and \sysname. From left to right, the optimized resource allocation and parallelization strategies of generation, training and inference are applied in order.}}}
    \vspace{-7mm}
    \label{fig:breakdown-mfc}
\end{figure}

\arxiv{
To illustrate the source of \sysname's performance improvement, we break down and analyze the time cost per training iteration across several representative experimental settings.
}

\paragraph{Function-call Level Breakdown.}
\arxiv{
First, we analyze the wall time of model function calls in two representative cases: a 7B actor with a 7B critic, and a 70B actor with a 7B critic. These cases respectively feature identical/similar and different sizes for the actor and critic models.
\Cref{fig:breakdown-mfc} shows the wall time per training step in \sysname with progressively applied optimizations:
\begin{itemize}[leftmargin=*]
\setlength{\itemsep}{0pt}
    \item CUDAGraph Generation.
    \item Generation parallelization
    \item Training parallelization \& concurrent execution.
    \item Inference parallelization \& concurrent execution. 
\end{itemize}
Performance improves incrementally from \sysname-Heuristic (leftmost bar) to \sysname (rightmost bar), with each step adding one optimization.
The orange and blue bars demonstrate the impact of CUDAGraph generation, a key contributor to performance improvement. 
The primary difference between the two settings lies in training phase optimization.
In the 7B+7B configuration, \sysname concurrently executes actor and critic training on separate devices. Since their training times are similar, this creates perfect overlap, maximizing performance. In contrast, for configurations with large model size disparities like 70B+7B, \sysname executes training sequentially on the global device mesh. The significant computational imbalance makes concurrent execution inefficient, so \sysname instead employs tailored parallelization strategies for each model to optimize overall performance.
The details of the execution plans and wall time breakdown are presented in \Cref{tab:parallel-strat-7-7-m,tab:parallel-strat-7-7-s,tab:parallel-strat-70-7-m,tab:parallel-strat-70-7-s,tab:wall-time-breakdown}.
}

\begin{figure}
    \centering
    \includegraphics[width=\columnwidth]{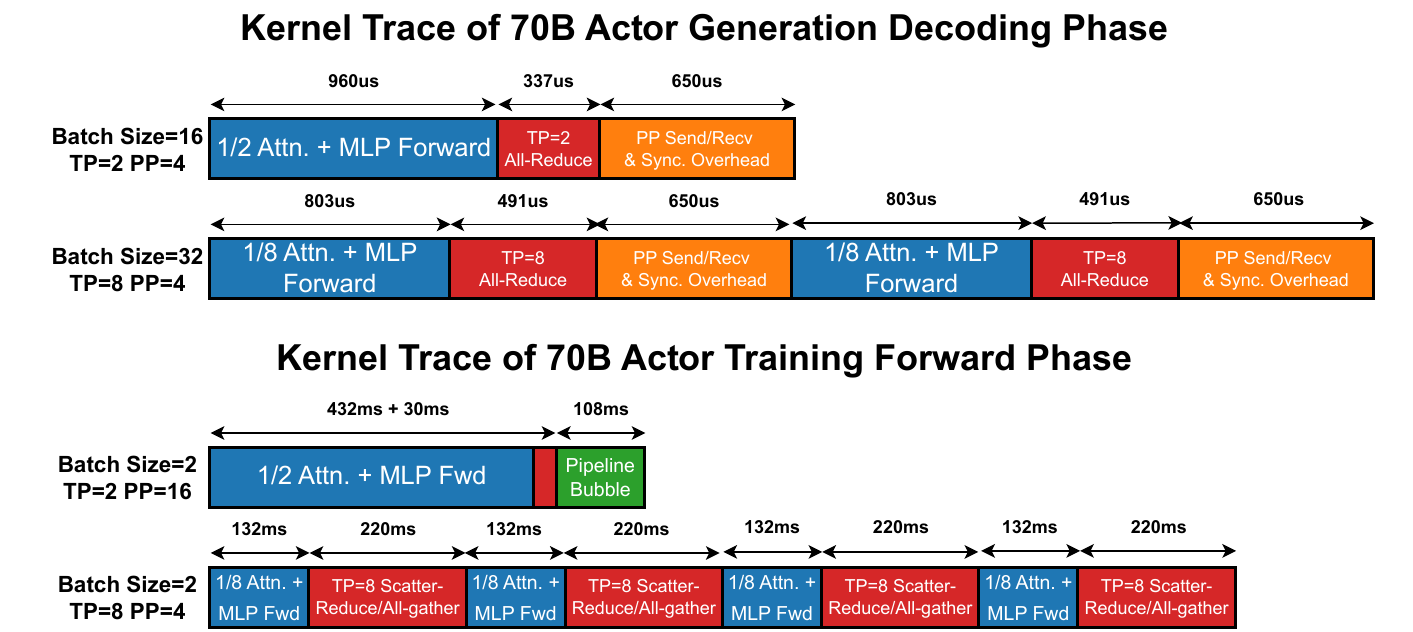}
    \caption{\small{Simplified kernel traces of a transformer layer completing the same amount of decoding or training computation. The top trace in each sub-figure represents \sysname, while the bottom trace represents \sysname-Heuristic. \sysname's parallelization strategies reduce memory I/O by invoking fewer computation kernels and minimize communication overhead caused by excessive tensor or pipeline parallelism.}}
    \vspace{-4mm}
    \label{fig:trace-example}
\end{figure}

\paragraph{Kernel Level Breakdown.}
To understand the enhancement of transforming parallelization strategies, we further examine the CUDA kernel traces, with a simplified example shown in \Cref{fig:trace-example}.
\arxiv{During decoding, \sysname prioritizes TP over PP to avoid the significant synchronization overhead between pipeline stages caused by numerous small decoding steps. Additionally, \sysname maximizes the DP degree within available GPU memory constraints. This reduces memory I/O and P2P communication overheads with less kernel invocations.}
For the compute-bounded training phase, \sysname utilizes a larger PP degree with a large number of micro-batches. Consequently, it minimizes the TP-induced collective communication overhead with a minimal bubble time increase.

\begin{figure}
    \centering
    \includegraphics[width=\columnwidth]{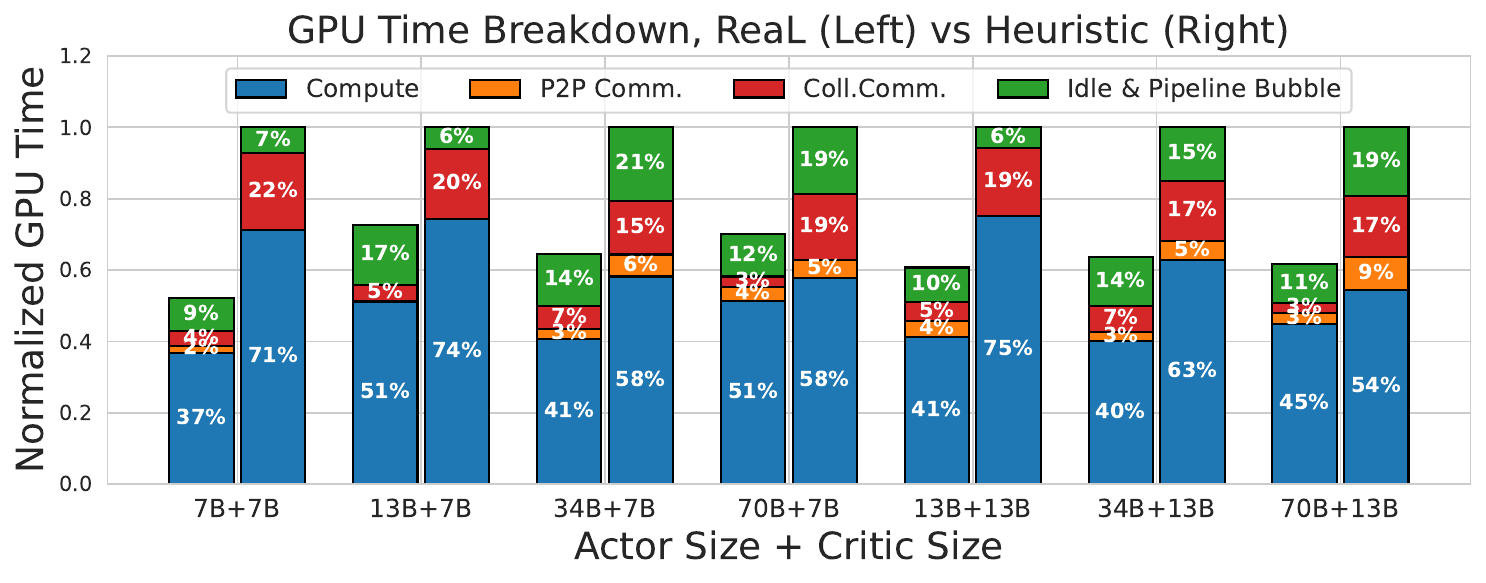}
    \vspace{-6mm}
    \caption{\small{
    The CUDA kernel time statistics of an RLHF iteration for \sysname (left) and \sysname-Heuristic (right). \sysname effectively eliminates the overhead of parallelization, i.e., the collective communication of TP and the P2P communication of PP, and reduces the memory IO time in compute kernels.
    }}
    \vspace{-5mm}
    \label{fig:breakdown}
\end{figure}

We further validate these observations by decomposing the GPU time per training iteration into three CUDA kernel types, as illustrated in \Cref{fig:breakdown}. 
\sysname demonstrates a similar trend of the kernel time decreasing across all scenarios.
We also note that the broadcasts of data transfer and parameter reallocation take much less GPU time than visualized types, so we omit them from the figure.
To conclude, the improvement of \sysname stem from two key aspects of our execution plan design. First, given a fixed device count, \sysname optimizes the parallelization strategies to minimize redundant memory IO and communication overheads from excessive TP or PP degrees. Second, by executing function calls concurrently across different device subsets, \sysname reduces per-function communication overheads through decreasing parallelization degrees.

\begin{figure}[t!]
    \centering
    \includegraphics[width=0.98\columnwidth]{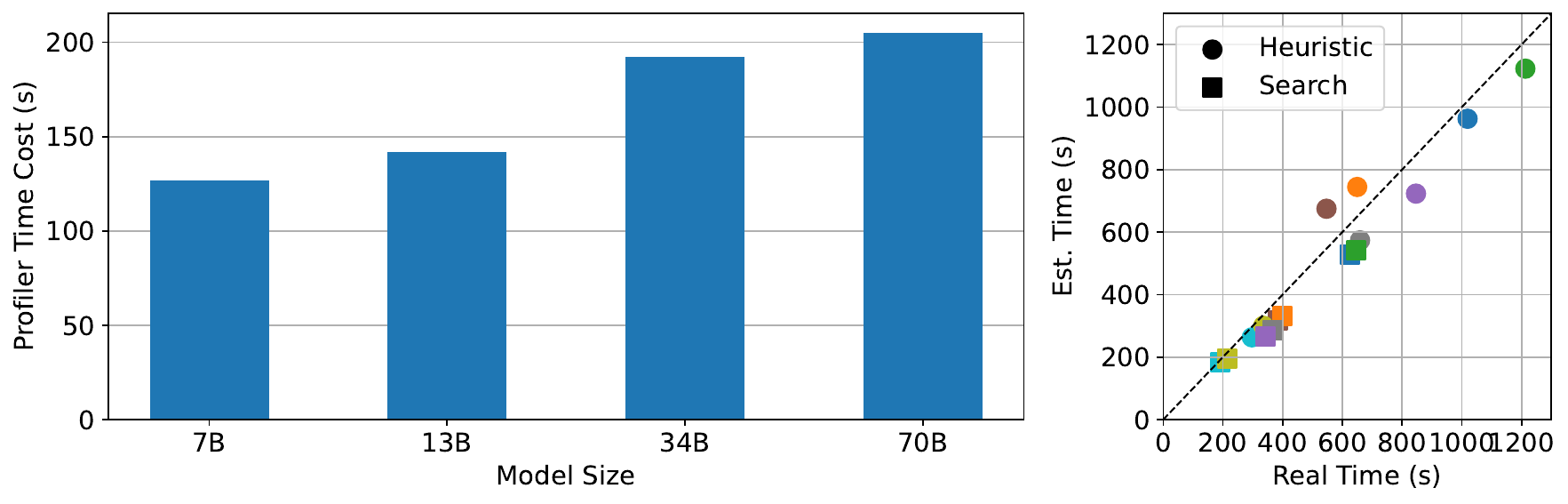}
    \vspace{-4mm}
    \caption{\small{
    (Left) The time of profiling before cost estimation.
    We consider batch sizes ranging from 1 to 512 and sequence lengths limited to 256, 512, and 1024. 
    (Right) The estimated time cost produced by the estimator and the real time cost of execution plans used in experiments.
    % that compare the throughput of \sysname and \sysname-Heuristic (\Cref{fig:heuristic-ctx-bs}). 
    % One data point denotes an execution plan. 
    Two data points of a same color denote the searched and heuristic execution plan in one experiment setting.
    }}
    \vspace{-3mm}
    \label{fig:profiler_error_rate}
\end{figure}

\begin{figure}[t!]
    \centering
    \includegraphics[width=0.98\columnwidth]{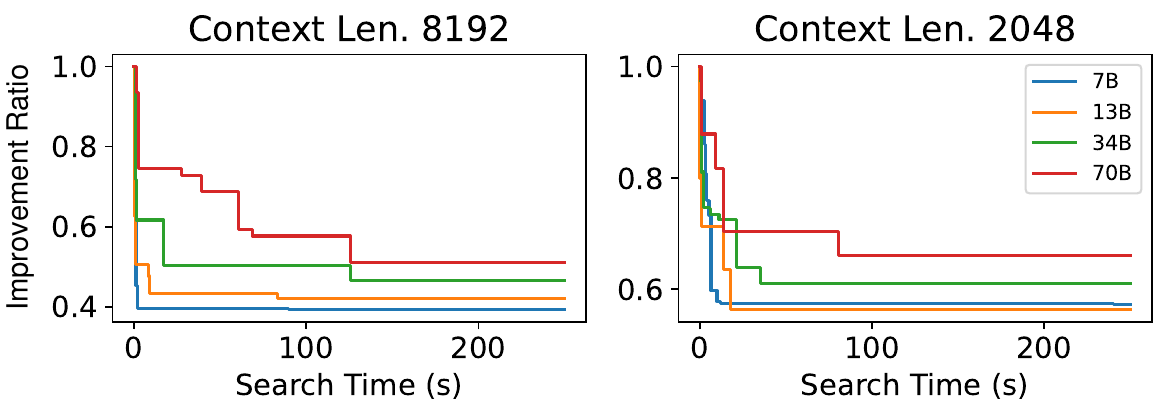}
    \vspace{-4mm}
    \caption{\small{
    The time cost of the best discovered execution plan compared to the initial one as the searching process proceeds. We name this metric as \textbf{improvement ratio}.  
    % The ratio of estimated time cost with the best searched and initial execution plans achieved by different search times. The search engine runs on a single CPU for 15 minutes. Parts of the lines after reaching a plateau are omitted.
    }}
    \vspace{-4mm}
    \label{fig:search_time}
    %\vspace{-4mm}
\end{figure}

\begin{figure}[t!]
    \centering
    \includegraphics[width=0.67\columnwidth]{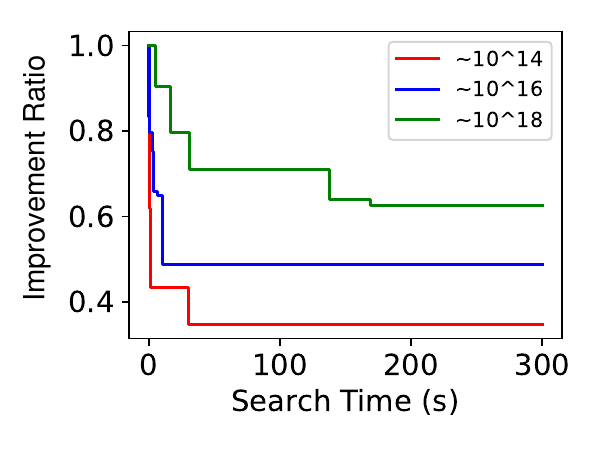} % Use \linewidth for consistency
    \vspace{-4mm}
    \caption{\arxiv{\small{The performance of the MCMC-based search algorithm with pruning in an experiment setting with 1024 GPUs. Three lines show performance with the search spaces that are pruned to $10^{14}$, $10^{16}$, and $10^{18}$ execution plans.}}}
    \label{fig:search_time_1024GPUs}
    \vspace{-4mm}
    %\vspace{-4mm}
\end{figure}

\begin{figure}[t!]
    \centering
    \includegraphics[width=0.75\columnwidth]{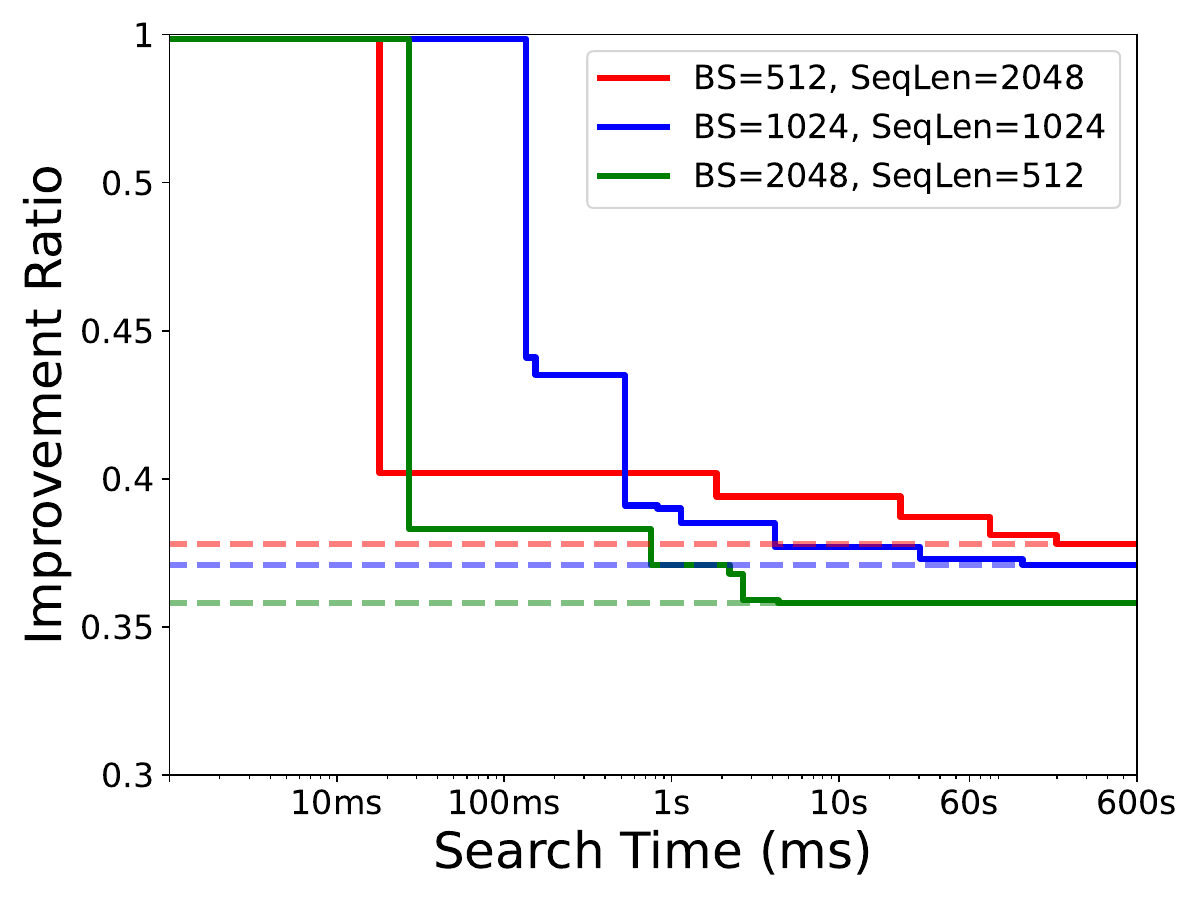} % Match width with first subfigure
    \vspace{-4mm}
    \caption{\arxiv{\small{The performance of execution plans produced by MCMC-based search in 10 minutes, in the setting of 7B+7B model sizes and 8 GPUs and three different settings of batch sizes and sequence lengths. The x-axis is log scaled, and the dotted lines mark the optimal performance produced by brute-force search.}}}
    \vspace{-4mm}
    \label{fig:bruteforce_comparison}
    %\vspace{-4mm}
\end{figure}

\iffalse
\begin{figure}[htbp]
    \centering % Ensures the entire figure is centered
    \begin{minipage}[t]{0.49\columnwidth}
        \centering
        \includegraphics[width=\linewidth]{documents/figures/1024GPU_search_ablation.pdf} % Use \linewidth for consistency
        \caption{\small The performance of the MCMC-based search algorithm with pruning in an experiment setting with 1024 GPUs. Three lines show performance with the search spaces that are pruned to $10^{14}$, $10^{16}$, and $10^{18}$ execution plans.}
        \label{fig:search_time_1024GPUs}
    \end{minipage}
    \hfill % Adds flexible space between subfigures
    \begin{minipage}[t]{0.48\columnwidth}
        \centering
        \includegraphics[width=\linewidth]{documents/figures/bruteforce_search_new.pdf} % Match width with first subfigure
        \caption{\small The performance of execution plans produced by MCMC-based search using 5 seconds and 10 minutes, in the setting of 7B+7B model sizes and 8 GPUs. The dotted lines mark the optimal performance produced by brute-force search.}
        \label{fig:bruteforce_comparison}
    \end{minipage}
    % \caption{Comparison of MCMC-based search performance under different settings.} % Overall figure caption
\end{figure}
\subsection{Ablations of the Execution Plan Generator}
\fi

We evaluate three key aspects of the execution plan generator in \sysname, including the time cost of the profiler, the accuracy of the runtime estimator, and the performance of the search engine. 

\textbf{Profiler.}
The profiler requires less than 4 minutes to collect a model's complete statistics, as shown in \Cref{fig:profiler_error_rate} (left). Our experiments only profile statistics of individual layers and inter/intra-node bandwidths. These profiled statistics are reusable across experiments within the same model family.

\textbf{Runtime Estimator Accuracy.}
The comparison between the real and estimated time costs, presented in \Cref{fig:profiler_error_rate} (right), shows relative differences consistently below 25\%. Crucially, the estimated costs maintain the same relative ordering as the real costs across different execution plans, ensuring the reliability of searched execution plans.

\textbf{Search Engine.}
\Cref{fig:search_time} tracks the estimated RLHF training cost throughout the searching process, using settings from the throughput experiments in \Cref{fig:heuristic-ctx-bs}. The search engine is able to identify execution plans with significant throughput improvements within 150 seconds across all experimental configurations. 

\arxiv{
As the number of GPUs increases, the search space grows at a high-degree polynomial rate. When the number of GPUs reaches over 1000, the entire search space has more than $10^{24}$ execution plans. In this case, the efficiency of MCMC sampling can degrade badly. 
% However, we could prune a large number of parallelization and allocation choices of model function calls that occupy only a small portion of GPUs in the cluster, since they will probably result in large GPU idle time or out-of-memory issues.
To address this, we employ an effective heuristic pruning technique that eliminates suboptimal execution plans likely to cause excessive GPU idle time, out-of-memory errors, or high communication overhead. For instance, we discard parallelization strategies where the tensor parallelization degree exceeds the number of GPUs per node, as these incur significant communication bottlenecks due to limited inter-node bandwidth. We also prune execution plans where the model function calls do not fully utilize the device mesh, as such configurations inevitably lead to GPU idle time and suboptimal performance.
}

\arxiv{
In \Cref{fig:search_time_1024GPUs}, we present an ablation study that shows the relationship between the size of pruned search space and the efficiency of MCMC sampling in an experiment setting with 1024 GPUs. The results show that our algorithm can still find a fast execution plan within 5 minutes by pruning the search space.
}
Furthermore, while our demonstration uses a single-threaded implementation, the search process can be further accelerated through a multi-core parallelization.

\arxiv{
\textbf{The Optimality of MCMC-based Search.} 
\Cref{fig:bruteforce_comparison} demonstrates a comparison between the execution plan produced by the MCMC-based search algorithm and the optimal ones produced by brute-force search, in the setting of 7B+7B model sizes and 8 GPUs. The result in \Cref{fig:bruteforce_comparison} shows that, in this setting, our search algorithm could achieve more than $95\%$ of the best performance in 5 seconds. Moreover, our search algorithm could produce the optimal execution plans within 10 minutes.
}

%%%%%%%%%%%%%%%%%%%%%%%%%%%%%%%%%%%%%%%%%%%%%%%%%%%%%%%%%%%
\subsection{RLHF Algorithms Beyond PPO}
\label{sec:exp-beyond-ppo}

\begin{figure}[t!]
    \centering
    \includegraphics[width=0.85\columnwidth]{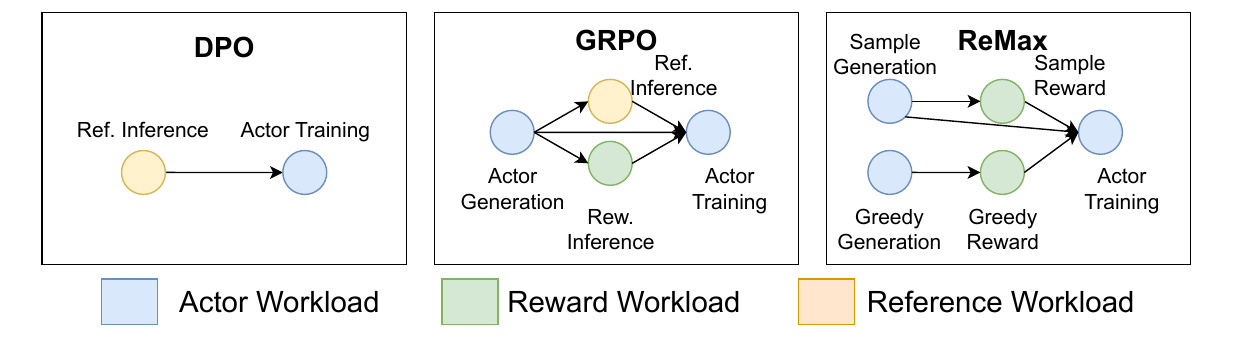}
    \vspace{0.5mm}
    
    \small
    \centering
    \begin{tabular}{c|ccc}
        \toprule
         \makecell{Throughput\\(PetaFLOP/s)} & \textbf{DPO} & \textbf{GRPO} & \textbf{ReMax} \\
        \midrule
        {ReaL-Heuristic} & 32.84 & 60.96 & 2.49 \\
        {ReaL (Ours)} & 50.38 & 71.12 & 7.23 \\
        \midrule
        {Improvement (\%)} & +53.4\% & +16.67\% & +190.84\% \\
        \bottomrule
    \end{tabular}
    \vspace{-2mm}
    \caption{\small{Throughput comparison with the heuristic execution plan on three prevalent RLHF algorithms other than PPO. Their dataflow graph representations are shown in the upper part.}}
    \vspace{-5mm}
    \label{fig:dpo-grpo-remax}
\end{figure}

\sysname can naturally incorporate any RLHF algorithms representable as a directed acyclic graph (DAG) with generation, inference, and training function calls.
We examine three concrete examples, including DPO~\cite{dpo}, GRPO~\cite{deepseekmath-grpo}, and ReMax~\cite{remax}.

In \Cref{fig:dpo-grpo-remax}, we compare \sysname with \sysname-Heuristic using a 70B Actor and 7B Critic on 16 nodes and observe an average throughput improvement of 87\%.
Among these three algorithms, ReMax achieves the highest gain by execution its two generation calls concurrently rather than sequentially. Conversely, GRPO shows more modest improvements due to its grouped generation technique. GRPO increases the batch size by 8× and makes the workload much more compute-bounded, which diminishes the benefits of reducing memory IO or TP/PP overheads.

%%%%%%%%%%%%%%%%%%%%%%%%%%%%%%%%%%%%%%%%%%%%%%%%%%%%%%%%%%%
\subsection{Strong Scaling Trend}

\begin{figure}
    \centering
    \includegraphics[width=\columnwidth]{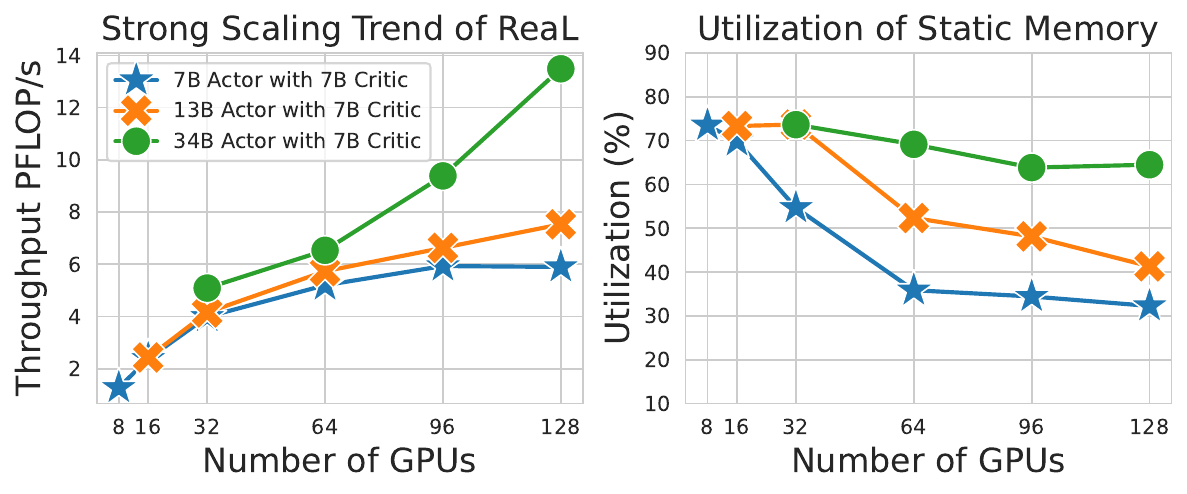}
    \vspace{-6mm}
    \caption{\small{The throughput and memory utilization in strong scaling experiments.
    \sysname can achieve (super-)linear scaling when the computation budget is tight by parallelizing computation and trading memory for communication. For a small model, the performance will hit the plateau due to the memory IO overhead of auto-regressive generation.}}
    \vspace{-4mm}
    \label{fig:strong-scaling}
\end{figure}

We analyze the \emph{strong scaling} performance by measuring throughput for fixed problem sizes across increasing device counts. \Cref{fig:strong-scaling} reveals a sub-linear scaling for 7B actors but a super-linear scaling for 34B actors.

\textbf{{Analysis.}}
The scaling behavior can be understood by examining the generation and training patterns, which dominate the RLHF iteration time. With limited computational resources, both operations are compute-bounded. Additional resources enable computation parallelization and can trade off more overall memory usage for less communication, leading to linear or super-linear gains. However, as resources increase, generation becomes a bottleneck due to the inherent sequential nature of autoregressive processing. It requires iterative KV cache loading, creating an irreducible overhead that leads to a diminishing scaling return.

\textbf{{Practical Suggestions.}}
Due to its larger algorithm design and hyperparameter searching space,
RLHF implementations face a fundamental trade-off between faster training with more resources and broader configuration exploration with fewer GPUs. The observed sub-linear scaling indicates the reduced resource efficiency at larger scales. However, a minimal resource allocation can impede production by extending experiment duration. Our results suggest identifying the transition point from super-linear to sub-linear scaling for each training configuration as the optimal device allocation. We recommend using static memory utilization as a heuristic metric, with utilization below 60\% indicating diminishing returns from additional GPU resources, as demonstrated in \Cref{fig:strong-scaling} (right).

\section{Related Work}

\subsection{Systems for Training and Serving LLMs}

Significant efforts have been made to develop distributed LLM training systems \cite{megatron2, palm, megascale} that leverage efficient data \cite{fsdp, zero}, tensor-model \cite{gshard, tofu}, and pipeline-model parallelism \cite{gpipe, terapipe}. Concurrently, research has focused on the efficient serving of pre-trained LLMs for generation \cite{flexgen, s-lora, orca, distserve}. However, integrating dependent workloads for training, inference, and generation, as in the case of RLHF, presents a challenge that extends beyond these individual efforts.

% Several concurrent works~\cite{adaptive-placement, puzzle-rlhf, hybridflow} explore RLHF system design and are conceptually similar to our paper.
% However, our formulation as an augmented dataflow graph, combined with a search-based execution plan generator and worker-based runtime engine, generalizes and advances previous solutions.
% Another concurrent work~\cite{rlhfuse} examines fusing different function calls in PPO, which a direct extension of our paper. Instead of optimizing or merging individual function calls, we propose a principled approach for efficiently executing a DAG composed of these calls. We identify parameter reallocation as the key to addressing this challenge, a factor overlooked by prior works.

\arxiv{Previous RLHF systems~\cite{deepspeedchat, openrlhf, nemo-aligner} typically employ hand-crafted parallelization strategies with limited flexibility. While they incorporate techniques like concurrent execution and ad hoc parameter resharding, these approaches remain inefficient and fail to adapt to diverse RLHF training scenarios. }
Several concurrent works~\cite{adaptive-placement, puzzle-rlhf, hybridflow} explore RLHF system design and are conceptually similar to our paper. 
In comparison, \sysname identifies parameter reallocation as the key to addressing this challenge, a factor overlooked by these works.
Moreover, our problem formulation and search-based solution offers more generalization and optimization opportunities.
Another concurrent work~\cite{rlhfuse} is an orthogonal extension to our paper. 
It proposes to fuse pipeline stages of actor and critic training and balancing the data skewness during generation, which targets on some special cases in the workflow of PPO.

\subsection{\small{GPU Memory Management for Distributed Training}}

Previous work on GPU memory management has primarily focused on reducing runtime memory usage, rather than improving training throughput. Techniques such as gradient checkpointing, ZeRO-3 optimization~\cite{zero}, and parameter offloading~\cite{zero-offload, zero-nvme, offload-training1, offload-training2} trade computation or communication to save memory. 
% We incorporate these methods to conserve GPU memory where feasible during the evaluation of \sysname and baselines.

Model parameter communication has been explored in parameter server architectures \cite{ps} and large-scale reinforcement learning systems \cite{openai5, srl}. These systems replicate the same set of parameters across multiple devices for concurrent job execution, with periodic synchronization for parameter updates. OpenRLHF~\cite{openrlhf} follows this pattern as well. Parameter synchronization is a specific case of parameter reallocation, where the source and destination devices are disjoint. However, in the context of LLMs, this technique often results in GPU underutilization, making it inefficient.

The concept most related to parameter reallocation is the HybridEngine in DSChat~\cite{dschat} and HybridFlow~\cite{hybridflow}. However, HybridEngine was limited to the actor model on the same device mesh. Parameter reallocation generalizes this approach, allowing it to be applied to any models within the algorithm, whether devices are disjoint or overlapping, leading to further throughput improvements, as demonstrated in~\Cref{tab:wall-time-breakdown}.

\subsection{Automatic Parallelization of DL Models}

Given the substantial effort required to manually design a parallelization strategy, numerous studies have focused on automating the parallelization of deep learning models~\cite{alpa, flexflow, dapple, pipedream, tofu}. Notably, Alpa~\cite{alpa} and FlexFlow~\cite{flexflow} offer general solutions for deep learning models that can be parsed into tensor operator graphs. Alpa leverages dynamic programming, while FlexFlow employs a custom search algorithm.

In theory, the entire RLHF training workflow could be represented as a tensor operation graph and automatically parallelized using prior methods. However, they are sub-optimal for two key reasons.
First, parameter reallocation introduces significant optimization opportunities in RLHF, but is unnecessary in supervised training. Consequently, previous methods do not consider parameter reallocation at runtime, resulting in subpar performance. Second, RLHF involves four different LLMs, which are highly operator-intensive. Searching through the entire tensor operator graph would be prohibitively expensive.
In contrast, \sysname accounts for parameter reallocation and operates at the granularity of model function calls. Our approach not only enhances end-to-end training performance but also explores a smaller solution space, significantly speeding up the search process.

% overcome this obstacle by operating on the granularity of model function calls, so the solution space is significantly reduced.
% , making it practical to search for optimized parallelization strategies of each model in RLHF. 

% \subsection{Task Scheduling on GPU Clusters}

% In RLHF, model function calls can be viewed as dependent tasks to be scheduled on a GPU cluster. While scheduling training jobs to a shared GPU cluster has been extensively studied in the literature \cite{schd1,schd2,schd3}, these works primarily address supervised training and struggle with generation/inference tasks involving large models and strong dependencies. Our paper shares a similar motivation to elastic training \cite{elast}, although existing works in this area predominantly target data parallel (DP) training and are inadequate for scaling model-parallelized LLMs.

% Conceptually, parameter reallocation resembles the GPU time-share mechanism in Gandiva \cite{gandiva}, which dynamically adjusts the set of GPUs used by a job through migration, expansion, or contraction.
% However, \sysname goes further by both changing device locations and parallelization strategies. Furthermore, \sysname leverages the predictability of RLHF function calls and offers the capability to automatically search for an execution plan before runtime.

\section{Conclusion}
\label{sec:conclusion}

In this paper, we present \sysname, the first system capable of automatically finding and executing a fast execution plan for RLHF training with parameter reallocation.  We evaluate the performance of \sysname against prior RLHF systems to demonstrate its superior performance. We believe that \sysname will not only democratize the powerful RLHF training algorithm but also encourage the development of novel algorithms on LLMs in the future.

\bibliography{example_paper}
\bibliographystyle{mlsys2025}

%%%%%%%%%%%%%%%%%%%%%%%%%%%%%%%%%%%%%%%%%%%%%%%%%%%%%%%%%%%%%%%%%%%%%%%%%%%%%%%
%%%%%%%%%%%%%%%%%%%%%%%%%%%%%%%%%%%%%%%%%%%%%%%%%%%%%%%%%%%%%%%%%%%%%%%%%%%%%%%
% SUPPLEMENTAL CONTENT AS APPENDIX AFTER REFERENCES
%%%%%%%%%%%%%%%%%%%%%%%%%%%%%%%%%%%%%%%%%%%%%%%%%%%%%%%%%%%%%%%%%%%%%%%%%%%%%%%
%%%%%%%%%%%%%%%%%%%%%%%%%%%%%%%%%%%%%%%%%%%%%%%%%%%%%%%%%%%%%%%%%%%%%%%%%%%%%%%
\appendix

\clearpage
\section{Experiment Details}
\label{app:exp-detail}

\begin{table*}[]
    \centering
    \begin{tabular}{lcccc}
    \toprule
    Identifier & 7B & 13B & 34B & 70B \\
    \midrule
    HiddenSize & 4096 & 5120 & 8192 & 8192 \\
    IntermediateSize & 14336 & 13824 & 22016 & 28672 \\
    NumLayers & 32 & 40 & 48 & 80 \\
    NumAttentionHeads & 32 & 40 & 64 & 64 \\
    NumKVHeads & 8 & 40 & 8 & 8 \\
    VocabSize & 128256 & 128256 & 128256 & 128256 \\
    MaxPositionEmbeddings & 8192 & 8192 & 8192 & 8192 \\
    TotalParamCount & 8030261248 & 14001525760 & 35321028608 & 70553706496 \\
    ParamCount w./o. Output Embedding & 7504924672 & 13344855040 & 34270355456 & 69503033344 \\
    \bottomrule
    \end{tabular}
    \caption{The LLaMA-3 model configurations used in experiments. Because critic models have a smaller output embedding layer than the actor (i.e., the output dimension is 1 for the critic), we use the embedding-less parameter count as the identifier.}
    \label{tab:llama-config}
\end{table*}

Our base setting is adopted from \citet{instrgpt}, which utilizes a global batch size of 512, context length 2048, 
and a maximum prompt length of 1024. The global batch is divided into 8 mini-batches for PPO training.

We emphasize that the prompt and generation length may vary for different models, datasets or tasks, algorithm implementation, and even during RLHF training. To eliminate this effect and perform a fair comparison, we synthesize random data with the maximum prompt length and terminate generation only after the maximum length is reached.

We create LLaMA models of four different sizes with their detailed configurations shown in~\Cref{tab:llama-config}.
For weak scaling experiments, we increase the model size and batch size proportionally to the number of devices.
In particular, for 16, 32, 64, and 128 GPUs, the model sizes are 7B, 13B, 34B, and 70B, and batch sizes are 512, 1024, 2048, 4096, respectively.
For experiments with a longer context length, we fix the number of tokens in the global batch. For instance, when the context length increases from 2048 to 8192, the global batch size decreases by a factor of 4.
In experiments for strong scaling and dditional RLHF algorithms, we adopt the base setting with 70B actor/reference models and 7B critic/reward models on 16 nodes.

We show the execution plans of wall time breakdown examples (\Cref{tab:wall-time-breakdown}) in \Cref{tab:parallel-strat-7-7-m,tab:parallel-strat-7-7-s,tab:parallel-strat-70-7-m,tab:parallel-strat-70-7-s}.

\begin{table*}[]
    \centering
    \small
    \begin{tabular}{llccccc}
    \toprule
 & DeviceMesh & TP & PP & DP & \#Micro-Batches & Time \\
\midrule
ActorGen & trainer[01-16] & 2 & 4 & 16 & 4 & 185.1 \\
RewInf & trainer[01-16] & 1 & 8 & 16 & 4 & 5.6 \\
RefInf & trainer[01-16] & 1 & 8 & 16 & 16 & 35.6 \\
CriticInf & trainer[01-16] & 1 & 8 & 16 & 16 & 5.6 \\
CriticTrain & trainer[01-16] & 8 & 4 & 4 & 2 & 20.8 \\
ActorTrain & trainer[01-16] & 2 & 16 & 4 & 2 & 108.0 \\
\bottomrule
\end{tabular}
    \caption{Device allocations and parallelization strategies for the 70B Actor and 7B critic searched case in \Cref{tab:wall-time-breakdown}.}
    \label{tab:parallel-strat-70-7-s}
\end{table*}

\begin{table*}[]
    \centering
    \small
    \begin{tabular}{llccccc}
    \toprule
     & DeviceMesh & TP & PP & DP & \#Micro-Batches & Time \\
    \midrule
    ActorGen & trainer[01-16] & 8 & 4 & 4 & 8 & 241.8 \\
    RewInf & trainer[01-16] & 8 & 4 & 4 & 8 & 12.6 \\
    RefInf & trainer[01-16] & 8 & 4 & 4 & 8 & 63.5 \\
    CriticInf & trainer[01-16] & 8 & 4 & 4 & 8 & 12.5 \\
    CriticTrain & trainer[01-16] & 8 & 4 & 4 & 8 & 35.7 \\
    ActorTrain & trainer[01-16] & 8 & 4 & 4 & 8 & 163.4 \\
    \bottomrule
    \end{tabular}
    \caption{Device allocations and parallelization strategies for the 70B Actor and 7B critic heuristic case in \Cref{tab:wall-time-breakdown}.}
    \label{tab:parallel-strat-70-7-m}
\end{table*}

\begin{table*}[]
    \centering
    \small
    \begin{tabular}{llccccc}
    \toprule
 & DeviceMesh & TP & PP & DP & \#Micro-Batches & Time \\
\midrule
ActorGen & trainer[01-02] & 2 & 2 & 4 & 1 & 16.3 \\
RewInf & trainer01 & 2 & 1 & 4 & 16 & 6.0 \\
RefInf & trainer02 & 1 & 2 & 4 & 16 & 8.0 \\
CriticInf & trainer[01-02] & 1 & 2 & 8 & 8 & 4.7 \\
CriticTrain & trainer02 & 4 & 2 & 1 & 2 & 28.1 \\
ActorTrain & trainer01 & 2 & 4 & 1 & 2 & 26.6 \\
\bottomrule
\end{tabular}
    \caption{Device allocations and parallelization strategies for the 7B Actor and 7B critic searched case in \Cref{tab:wall-time-breakdown}.}
    \label{tab:parallel-strat-7-7-s}
\end{table*}

\begin{table*}[]
    \centering
    \small
    \begin{tabular}{llccccc}
    \toprule
 & DeviceMesh & TP & PP & DP & \#Micro-Batches & Time \\
\midrule
ActorGen & trainer[01-02] & 8 & 1 & 2 & 4 & 44.2 \\
RewInf & trainer[01-02] & 8 & 1 & 2 & 4 & 7.3 \\
RefInf & trainer[01-02] & 8 & 1 & 2 & 4 & 7.6 \\
CriticInf & trainer[01-02] & 8 & 1 & 2 & 4 & 6.8 \\
CriticTrain & trainer[01-02] & 8 & 1 & 2 & 4 & 24.3 \\
ActorTrain & trainer[01-02] & 8 & 1 & 2 & 4 & 24.7 \\
\bottomrule
\end{tabular}
    \caption{Device allocations and parallelization strategies for the 7B Actor and 7B critic heuristic case in \Cref{tab:wall-time-breakdown}.}
    \label{tab:parallel-strat-7-7-m}
\end{table*}

\begin{table*}
    \footnotesize
    \centering
    \begin{tabular}{c|cc|cc}
        \toprule
        \multirow{2}{*}{\makecell{Time (s)}}& \multicolumn{2}{c}{7B + 7B} & \multicolumn{2}{c}{70B + 7B} \\
         & \sysname & Heuristic & \sysname & Heuristic \\
        \midrule
        \makecell{ActorGen\\(with CUDAGraph)} & 16.3 & 44.2 & 185.1 & 241.8 \\
        \makecell{ActorGen\\(w.o. CUDAGraph)} & 34.5 & 104.6 & 185.1 & 241.8 \\
        RewInf & 6.0 & 7.3 & 5.6 & 12.6 \\
        RefInf & 8.0 & 7.6 & 35.6 & 63.5 \\
        CriticInf & 4.7 & 6.8 & 5.6 & 12.5 \\
        CriticTrain & 28.1 & 24.3 & 20.8 & 35.7 \\
        ActorTrain & 26.6 & 24.7 & 108.0 & 163.4 \\
        \midrule
        \makecell{End2End\\(with CUDAGraph)} & 64.0 & 122.6 & 383.1 & 546.8 \\
        \makecell{End2End\\(w.o CUDAGraph)} & 82.2 & 183.0 & 547.4 & 912.3 \\
        \bottomrule
    \end{tabular}
    \caption{The RLHF wall time breakdown of two most common and representative cases. \sysname reduces the end-to-end time by accelerating individual model function calls as well as concurrently executing independent computations.}
    \label{tab:wall-time-breakdown}
\end{table*}

\section{The API of \sysname}
\label{app:apis}
\begin{figure}[t!]
    \centering
    \includegraphics[width=0.9\columnwidth]{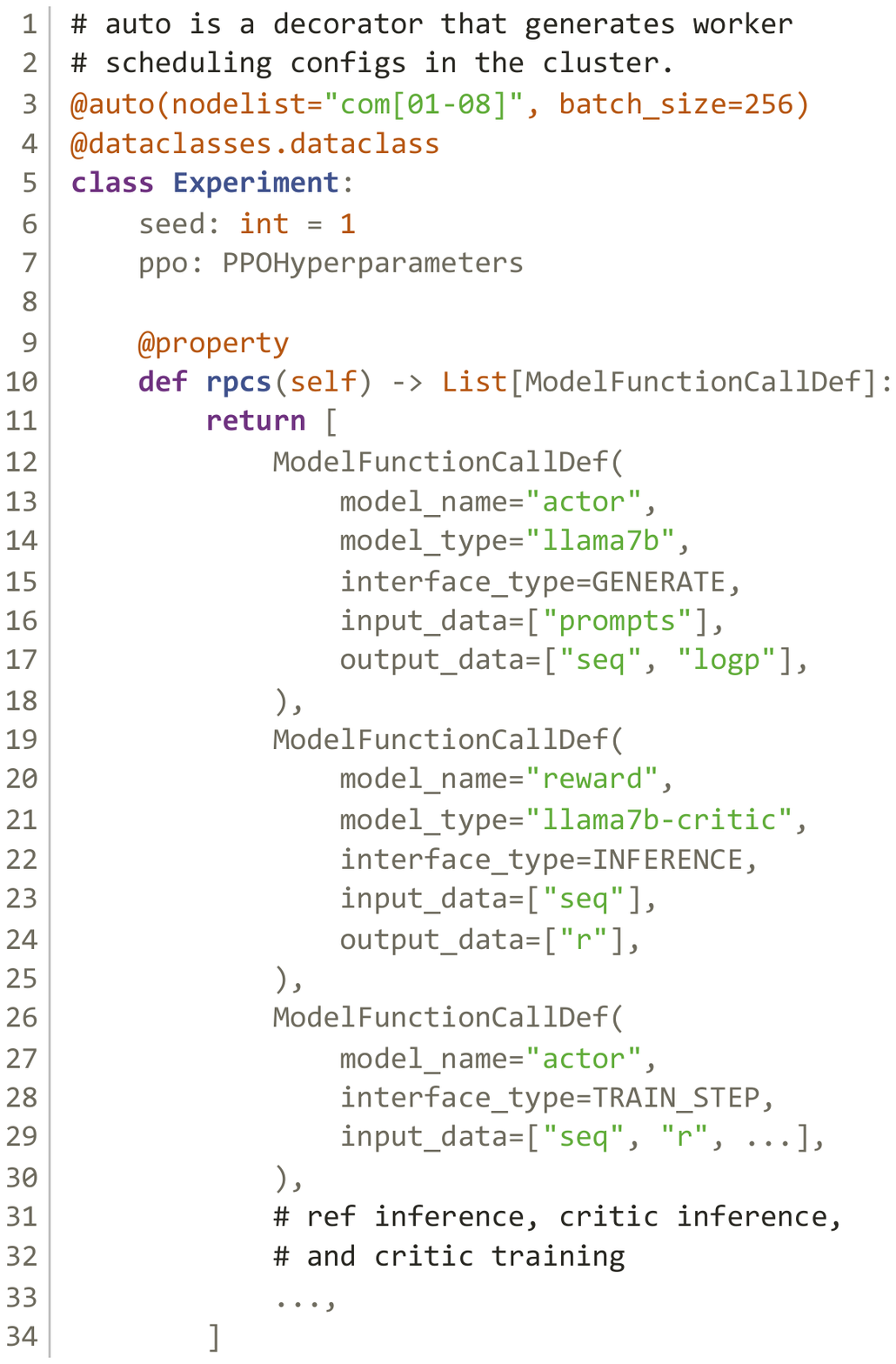}
    \caption{\small{An example of the user interface of \sysname. Given the dataflow graph (represented by a list of \texttt{ModelFunctionCallDef} objects), the training batch size, and cluster specifications, \sysname will automatically derive an execution plan via the \texttt{auto} decorator.}}
    \label{fig:api}
\end{figure}

% \paragraph{API}
% The API for an \sysname experiment is exemplified in \Cref{fig:api}. 
\Cref{fig:api} shows an example of the API for an \sysname experiment.
Users define the dataflow graph of the algorithm (e.g., RLHF) using a list of \texttt{ModelFunctionCallDef} objects. These objects encapsulate the model configuration and the function call type, along with specifying input and output data dependencies. Models sharing the same \texttt{model\_name} must have identical architectures (e.g., \texttt{llama7b}). 
They form parameter version dependencies, such that the inference and generation must wait for the training in the previous iteration.
The experiment configuration is then wrapped by the \texttt{auto} decorator, which initiates the search engine to derive an efficient execution plan. This plan is transformed into a scheduling configuration for launching workers, each assigned to a specific GPU or CPU via SLURM~\cite{slurm}.
The search engine and launcher both run under the hood.
Users are free to provide distinct interface implementations to implement a diverse range of training workflows.

\section{Simulation Algorithm}
\label{app:simulation_algo}

The simulation algorithm in show in \Cref{algo:simulate}.

\begin{algorithm}[ht]
\footnotesize
\begin{algorithmic}
\REQUIRE{The augmented dataflow graph $\mathcal{G}_p=(\mathcal{V}_p, \mathcal{E}_p)$, 
device meshes $D\in \mathcal{D}$ where $\mathcal{D}$ contains all valid device meshes in the cluster.}
% \KwResult{Output result}
% // Initialize attributes for nodes and device meshes\;
% \For{$v \in \mathcal{V}_p$}{
%    $v.startTime = 0$\;
%    $v.endTime = 0$\;
%    $v.readyTime = 0$\;
%}
%\For{$d$ \in \mathcal{D}}{
%    $d.last = None$
%}
\STATE ready\_queue = PriorityQueue()\textit{// Sorted by v.ReadyTime}\;
\STATE completed\_set = Set() \textit{// Contains completed nodes}\;    
\FOR{$v \in \mathcal{V}_p$}
\IF{$v.$parents=$\emptyset$}
\STATE ready\_queue.push($v$)
\ENDIF
\ENDFOR
\WHILE{\textnormal{!ready\_queue.empty()}}
    % \textit{// Execute ready node $v$}\;
\STATE Node $v$ = ready\_queue.pop()\;
\STATE DeviceMesh $D$ = $v$.device\_mesh\;
\STATE \textit{// $D$.last record the last completed node from all devices within $D$}\;
\STATE $v$.StartTime = max\{$v$.ReadyTime, $D$.last.EndTime\}\;
\STATE $v$.EndTime = $v$.StartTime + TimeCost($v$)\;
\STATE completed\_set.add($v$)\;    
\FOR{$D^\prime \in \mathcal{D}$}
\IF{\textnormal{overlap($D$, $D^\prime$) \textbf{and} $D^{\prime}$.last.EndTime $\le$ $D$.last.EndTime}}
\STATE $D^{\prime}$.last = $v$\;
\ENDIF
\ENDFOR
\FOR{$u \in v$.children} 
\STATE $u$.ReadyTime = max\{$u$.ReadyTime, $v$.EndTime\} \;
\IF{$w\in \textnormal{completed\_set}$ \textbf{for all} $w \in u$.parents}
\STATE ready\_queue.push($u$)\;
\ENDIF
\ENDFOR
\ENDWHILE
\STATE \textbf{return} \textnormal{max\{$v$.EndTime $| v\in \mathcal{V}_p$\}}
\end{algorithmic}
\caption{Calculate TimeCost($\mathcal{G}_p$)}
\label{algo:simulate}
\end{algorithm}

\section{Baselines}
\label{app:additional-baseline}

\arxiv{In \Cref{fig:other-sys-thpt}, we show the performance comparison between \sysname and 4 baseline RLHF systems: DeepSpeedChat~\cite{dschat}, OpenRLHF~\cite{openrlhf}, NeMoAligner~\cite{nemo-aligner} and veRL (HybridFlow~\cite{hybridflow}). 
The first three baselines are previous works of \sysname, and veRL is concurrent to \sysname.
In this section, we will briefly introduce the implementation of these baseline systems. We also list the version and backend of baseline systems used in our experiments.}

\arxiv{
DeepSpeedChat is developed using modules from a popular training backend DeepSpeed~\cite{deepspeed}. 
It supports sequential execution of model function calls, and uses TP for the generation task, ZeRO-3 DP for the training and inference task. 
It also implements HybridEngine, a technique that reshards parameters between actor training and generation.
}

\arxiv{
OpenRLHF exploits vLLM~\cite{vllm} as their generation backend and DeepSpeed ZeRO-3 DP as their training backend. 
It divides GPUs into three groups, holding the actor/reference model, the critic/reward model and the vLLM generation engine separately. 
It allows the concurrent execution of actor and critic training. However, the generation and training phase can not be executed concurrently due to data and parameter dependencies. 
This results in a significant GPU idle time.
}

\arxiv{
Similarly, NeMoAligner divides GPUs into 2 disjoint GPU groups. 
Unlike OpenRLHF, it locates actor training and generation on the same GPU group. 
It splits the computations into micro batches and pipeline them to reduce the GPU idle time.
It exploits TRT-LLM~\cite{trtllm} (supports TP and resharding) as generation backend and Megatron-LM~\cite{megatron} as training backend (supports 3D parallelization).
}

\arxiv{
veRL supports colocating models on GPUs and split placement of models on different GPU groups, including the strategies adopted by three previous systems. 
It provides different choices for the generation (SGLang~\cite{sglang} and vLLM~\cite{vllm}) and training backend (Megatron-LM~\cite{megatron} and Pytorch FSDP~\cite{pytorch-fsdp}) to support different parallelization strategies. 
}

\arxiv{
We list the version and backend of baseline systems used in our experiments in \Cref{tab:version}. 
We remark that in this experiment, \sysname uses its own generation backend, model and pipeline parallelization, and adopts tensor parallelization and optimizer implementation from Megatron-LM. 
In a more recent version of \sysname, we also support vLLM and SGLang as generation backend, which is not included in the experiments in this paper. 
}

\begin{table*}[t]
    \footnotesize
    \centering
    \begin{tabular}{c|c|c|c}
        \toprule
          System & Version & Generation Backend & Training Backend \\
        \midrule
          DeepSpeedChat & commit f73a6ed & DeepSpeed v0.15.1 & DeepSpeed v0.15.1 \\
          OpenRLHF & v0.4.2 & vLLM v0.4.2 & DeepSpeed v0.15.0 \\
          NeMoAligner & v0.4.0 & TRT-LLM v0.10.0 & Megatron-LM v0.8.0 \\
          veRL & 0.2.0.post2 & vLLM v0.6.3 & Pytorch FSDP v2.4.0 \\
        \bottomrule
    \end{tabular}
    \caption{The version, generation backend and training backend used in our baseline experiments.}
    \label{tab:version}
\end{table*}

%%%%%%%%%%%%%%%%%%%%%%%%%%%%%%%%%%%%%%%%%%%%%%%%%%%%%%%%%%%%%%%%%%%%%%%%%%%%%%%
%%%%%%%%%%%%%%%%%%%%%%%%%%%%%%%%%%%%%%%%%%%%%%%%%%%%%%%%%%%%%%%%%%%%%%%%%%%%%%%

\end{document}